\documentclass[11pt]{article}

\pdfoutput=1

\usepackage[backref,linktocpage,bookmarks,plainpages=false]{hyperref}
\usepackage{amsmath}
\usepackage{amsthm}
\usepackage{amsfonts}          
\usepackage{amssymb}           
\usepackage{color}
\usepackage{graphicx}
\usepackage{cite}
\usepackage{multirow}

\numberwithin{equation}{section}

\usepackage[isu,bf]{caption}
\newlength{\xtrawidth}
\newlength{\xtraheight}
\renewcommand{\(}{\left(}
\renewcommand{\)}{\right)}
\newcommand{\Z}{\ensuremath{\mathbb{Z}}}

\newcommand{\R}{\ensuremath{{\mathbb{R}}}}
\newcommand{\C}{\ensuremath{{\mathbb{C}}}}
\newcommand{\Rep}[1]{\ensuremath{\mathbf{\underline{#1}}}}

\newcommand{\Kcone}{\ensuremath{\mathcal{K}}}
\newcommand{\tmod}{~\mathrm{mod}~}
\newcommand{\Ocal}{\ensuremath{{\cal O}}}
\newcommand{\Vvis}{\ensuremath{V^{(1)}}}
\newcommand{\Vhid}{\ensuremath{V^{(2)}}}
\newcommand{\Rhat}{\ensuremath{{\widehat R}}}

\DeclareMathOperator{\tr}{tr}
\DeclareMathOperator{\Span}{span}
\DeclareMathOperator{\rank}{rank}
\DeclareMathOperator{\re}{Re}

\renewcommand{\thefootnote}{\fnsymbol{footnote}}

\begin{document}

\begin{titlepage}
\title{\huge Vacuum Constraints for Realistic Strongly Coupled Heterotic M-Theories{\LARGE \\[.5cm]  }}
                       
\author{\Large{Burt A.~Ovrut {{*}}} \\[8mm]
     {Department of Physics, University of Pennsylvania} \\
     {Philadelphia, PA 19104--6396}\\[3mm] }

\date{}

\maketitle

\begin{abstract}

The compactification from the 11-dimensional Horava-Witten orbifold to 5-dimensional heterotic M-theory on a Schoen Calabi-Yau threefold is reviewed, as is the specific $SU(4)$ vector bundle leading to the ``heterotic standard model'' in the observable sector. A generic formalism for a  consistent hidden sector gauge bundle, within the context of strongly coupled heterotic M-theory, is presented. Anomaly cancellation and the associated bulk space 5-branes are discussed in this context. The further compactification to a 4-dimensional effective field theory on a linearized BPS double domain wall is then presented to order $\kappa_{11}^{4/3}$. Specifically, the generic constraints required for anomaly cancellation and by the linearized domain wall solution, as well as restrictions imposed by the  necessity to have  positive squared gauge couplings to order $\kappa_{11}^{4/3}$ are presented in detail.  Finally, the expression for the Fayet-Iliopoulos term associated with any anomalous $U(1)$ gauge connection is presented and its role in $N=1$ spontaneous supersymmetry breaking in the low energy effective theory is discussed.

\noindent

\let\thefootnote\relax\footnotetext{* Submitted to the special issue "Supersymmetric Field Theory 2018" in {\it Symmetry}}
\let\thefootnote\relax\footnotetext{~~~ovrut@elcapitan.hep.upenn.edu}

\vspace{.3in}
\noindent
\end{abstract}

\thispagestyle{empty}
\end{titlepage}

\tableofcontents

\section{Introduction}

One of the major prerogatives of the Large Hadron Collider (LHC) at CERN is to search for low-energy $N=1$ supersymmetry. It has long been known that specific vacua of both the weakly coupled \cite{Gross:1985fr,Gross:1985rr} and strongly coupled \cite{Witten:1996mz,Horava:1995qa,Horava:1996ma} $E_{8} \times E_{8}$  heterotic superstring can produce effective theories with at least a quasi-realistic particle spectrum exhibiting $N=1$ supersymmetry \cite{Greene:1986ar,Greene:1986bm,Greene:1986jb,Matsuoka:1986vg,Greene:1987xh}. It has also been shown that the various moduli associated with these theories can, in principle, be stabilized \cite{Anderson:2011cza}.
It is of considerable interest, therefore, to examine these low-energy theories in detail and to confront them, and their predictions, with present and up-coming CERN data. 

There are several important criterion that such theories should possess. First, they must exhibit a low-energy spectrum in the observable sector that is not immediately in conflict with known phenomenology. In this paper, we will choose the so-called ``heterotic standard model'' introduced in \cite{Braun:2005nv,Braun:2004xv,Braun:2005zv,Braun:2006ae,Braun:2006me}. The observable sector of this theory, associated with the first $E_{8}$ gauge factor, contains exactly the matter spectrum of the minimal supersymmetric standard model (MSSM) augmented by three right-handed neutrino chiral superfields, one per family, and a single pair of Higgs-Higgs conjugate superfields. There are no vector-like pairs of superfields or exotic matter of any kind. The low-energy gauge group of this theory is the $SU(3)_{C} \times SU(2)_{L} \times U(1)_{Y}$ of the standard model enhanced by a single additional gauged $U(1)$ symmetry, which can be associated with $B-L$. Interestingly, the observable sector of this model--derived from the ``top down'' in the $E_{8} \times E_{8}$  heterotic superstring--is identical to the minimal, anomaly free $B-L$ extended MSSM, derived from the ``bottom up'' in \cite{Barger:2008wn,FileviezPerez:2009gr}. The requisite radiative breaking of both the $B-L$ and electroweak symmetry, the associated mass hierarchy \cite{Ambroso:2009jd,Ambroso:2009sc,Ambroso:2010pe,Ovrut:2012wg,Ovrut:2015uea}, and various phenomenological \cite{FileviezPerez:2009bw,FileviezPerez:2012mj} and cosmological \cite{Brelidze:2010hf,Battarra:2014tga,Koehn:2013upa,Koehn:2012ar} aspects of this low-energy theory have been discussed. Their detailed predictions for LHC observations are currently being explored \cite{Marshall:2014kea,Marshall:2014cwa,Dumitru:2018jyb,Dumitru:2018nct,Dumitru:2019cgf}.
The reader is also referred to works on heterotic standard models in \cite{Candelas:2007ac,Anderson:2011ns,Anderson:2012yf,Anderson:2011vy}

The second important criterion is that there be a hidden sector, associated with the second $E_{8}$ gauge factor, that, prior to its spontaneous breaking, exhibits $N=1$ supersymmetry. The original papers on the heterotic standard model, in analogy with the KKLT mechanism \cite{Kachru:2003aw} in Type II string theory, allowed the hidden sector to contain an anti-5-brane--thus explicitly breaking $N=1$ supersymmetry \cite{Gray:2007zza,Gray:2007qy}. This was done so that the potential energy could admit a meta-stable de Sitter space vacuum. Such a vacuum was shown to exist, and its physical properties explored, in \cite{Braun:2006th}. Be this as it may, it is important to know whether or not completely $N=1$ supersymmetric hidden sectors can be constructed in this context. It was demonstrated in \cite{Braun:2006ae} that the hidden sector of the heterotic standard model satisfies the Bogomolov inequality, a non-trivial necessary condition for supersymmetry \cite{MR522939} (see also \cite{Douglas:2006jp,Andreas:2010hv,Andreas:2011zs}). More recently, within {\it weakly} coupled $E_{8} \times E_{8}$ heterotic superstring theory, an $N=1$ supersymmetric hidden sector was explicitly constructed using holomorphic line bundles \cite{Braun:2013wr}. This vacuum satisfies all conditions required to have a consistent compactification in the weakly coupled context--and is an explicit proof that completely $N=1$ supersymmetric heterotic standard models exist. Be that as it may, it remains important to construct a completely $N=1$ supersymmetric vacuum state of {\it strongly} coupled heterotic string theory that is totally consistent with all physical requirements. In this paper, we present the precise constraints required by such a vacuum within the context of strongly coupled heterotic M-theory.

Specifically, we do the following. In Section 2, the focus is on the compactification from the $11$-dimensional Horava-Witten  orbifold to $5$-dimensional heterotic M-theory--presenting the relevant details of the explicit Calabi-Yau threefold
and the observable sector $SU(4)$ vector bundle of the heterotic standard model. Following a formalism first presented in \cite{Blumenhagen:2005ga,Blumenhagen:2005zg,Blumenhagen:2006ux,Weigand:2006yj,Blumenhagen:2006wj,Blumenhagen:2005pm}, we discuss the generic structure of a large class of the hidden sector gauge bundles--a Whitney sum of a non-Abelian $SU(N)$ bundle and line bundles. Bulk space 5-branes \cite{Lukas:1998hk,Donagi:1999jp,Lukas:1999kt} and the anomaly cancellation constraint are presented in this context. The compactification of $5$-dimensional heterotic M-theory to the $4$-dimensional low-energy effective field theory on a BPS linearized double domain wall is then described. In particular, we derive the constraints that must be satisfied in order for the linearized approximation to be valid. Section 3 is devoted to discussing the $4$-dimensional $E_{8} \times E_{8}$ effective theory; first the lowest order $\kappa_{11}^{2/3}$ Lagrangian \cite{Lukas:1997fg}--along with the slope-stability criteria for the observable sector $SU(4)$ vector bundle--and then the exact form of the order $\kappa_{11}^{4/3}$ corrections \cite{Anderson:2009nt}. These corrections for both the observable and hidden sector gauge couplings and the Fayet-Iliopoulos terms are presented in a unified formalism. They are shown to be of the same form as in the weakly coupled string with the weak coupling constants replaced by a moduli-dependent expansion parameter. The constraints required for there to be positive squared gauge coupling parameters, as well as the constraints so that the low energy theory be either $N=1$ supersymmetric or to exhibit spontaneously broken $N=1$ supersymmetry, are then specified. Finally, we consider a specific set of vacua for which the hidden sector vector bundle is restricted to be the Whitney sum of one non-Abelian $SU(N)$ bundle with a single line bundle, while allowing only one five-brane in the bulk space. Under these circumstances, the constraint equations greatly simplify and are explicitly presented. We demonstrate how the parameters associated with these bundles are computed for a specific choice of these objects.

\section{The Compactification Vacuum}

$N=1$ supersymmetric heterotic M-theory on four-dimensional Minkowski
space $M_4$ is obtained from eleven-dimensional Horava-Witten theory
via two sequential dimensional reductions: First with respect to a
Calabi-Yau threefold $X$ whose radius is assumed to be smaller than
that of the $S^1/\Z_2$ orbifold and second on a ``linearized'' BPS
double domain wall solution of the effective five-dimensional
theory. Let us present the relevant information for each of these
within the context of the heterotic standard model.

\subsection{The d=11$\rightarrow$d=5 Compactification}

\subsubsection{The Calabi-Yau Threefold} 

The Calabi-Yau manifold $X$ is chosen to be a torus-fibered threefold
with fundamental group $\pi_1(X)=\Z_3 \times \Z_3$. Specifically, it
is a fiber product of two rational elliptic $d\mathbb{P}_{9}$
surfaces, that is, a self-mirror Schoen threefold~\cite{MR923487,Braun:2004xv}, quotiented with respect to a freely acting $\Z_3
\times \Z_3$ isometry. Its Hodge data is $h^{1,1}=h^{1,2}=3$ and,
hence, there are three K\"ahler and three complex structure
moduli. The complex structure moduli will play no role in the present
paper. Relevant here is the degree-two Dolbeault cohomology group
\begin{equation}
  H^{1,1}\big(X,\C\big)=
  \Span_\C \{ \omega_1,\omega_2,\omega_3 \} 
  \label{1}
\end{equation}
where $\omega_i=\omega_{ia {\bar{b}}}$ are harmonic $(1,1)$-forms on
$X$ with the property
\begin{equation}
  \omega_3\wedge\omega_3=0 ,\quad
  \omega_1\wedge\omega_3=3~\omega_1\wedge\omega_1 ,\quad
  \omega_2\wedge\omega_3=3~\omega_2\wedge\omega_2 \ .
  \label{2}
\end{equation}
Defining the intersection numbers as
\begin{equation}
  d_{ijk} = \frac{1}{v}
  \int_X \omega_i \wedge \omega_j \wedge \omega_k
  , \quad i,j,k=1,2,3
  \label{3}
\end{equation}
where $v$ is a reference volume of dimension $(\text{\it length})^6$,
it follows from \eqref{2} that
\begin{equation}\label{4}
  (d_{ijk}) = 
  \left(
    \begin{array}{ccc}
      \(0,\tfrac13,0\) & \(\tfrac13,\tfrac13,1\) & \(0,1,0\) \\
      \(\tfrac13,\tfrac13,1\) & \(\tfrac13,0,0\) & \(1,0,0\) \\
      \(0,1,0\) & \(1,0,0\) & \(0,0,0\)
    \end{array} 
  \right) .
\end{equation}
The $(i,j)$-th entry in the matrix corresponds to the triplet
$(d_{ijk})_{k=1,2,3}$.

Our analysis will require the Chern classes of the tangent bundle
$TX$. Noting that the associated structure group is $SU(3) \subset
SO(6)$, it follows that $\rank(TX)=3$ and $c_1(TX)=0$. Furthermore,
the self-mirror property of this specific threefold implies
$c_3(TX)=0$. Finally, we find that
\begin{equation}
  c_2(TX) = \frac{1}{v^{2/3}}
  \big(12 \omega_1\wedge\omega_1+12\omega_2\wedge\omega_2) .
  \label{5}
\end{equation}
We will use the fact that if one chooses the generators of $SU(3)$ to
be hermitian, then the second Chern class of the tangent bundle can be
written as
\begin{equation}
  c_2(TX)=-\frac{1}{16\pi^2} \tr_{SO(6)} R \wedge R ,
  \label{6}
\end{equation}
where $R$ is the Lie algebra valued curvature two-form.

Note that $H^{2,0}=H^{0,2}=0$ on a Calabi-Yau threefold. It follows
that $H^{1,1}(X,\C)=H^2(X,\R)$ and, hence, $\omega_i$, $i=1,2,3$ span
the real vector space $H^2(X,\R)$. Furthermore, it was shown
in~\cite{Braun:2005zv} that the curve Poincare dual to each two-form $\omega_i$ is
effective. It follows that the K\"ahler cone is the positive octant
\begin{equation}
  \Kcone = H^2_{+}(X,\R)
  \subset H^2(X,\R) .
\label{7}
\end{equation}
The K\"ahler form, defined to be $\omega_{a {\bar{b}}}=ig_{a
  {\bar{b}}}$ where $g_{a {\bar{b}}}$ is the Calabi-Yau metric, can be
any element of $\Kcone$. That is, suppressing the Calabi-Yau indices,
the K\"ahler form can be expanded as
\begin{equation}
  \omega = a^i\omega_i 
  , \quad a^i >0, \quad i=1,2,3 .
\label{8}
\end{equation}
The real, positive coefficients $a^i$ are the three $(1,1)$ K\"ahler
moduli of the Calabi-Yau threefold. Here, and throughout this paper,
upper and lower $H^{1,1}$ indices are summed unless otherwise
stated. The dimensionless volume modulus is defined by
\begin{equation}
  V=\frac{1}{v} \int_X \sqrt[6]{g}
  \label{9}
\end{equation}
and, hence, the dimensionful Calabi-Yau volume is ${\bf{V}}=vV$. Using the
definition of the K\"ahler form and \eqref{3}, $V$ can be written as
\begin{equation}
  V=\frac{1}{6v}\int_X
  \omega \wedge \omega \wedge \omega=
  \frac{1}{6} d_{ijk} a^i a^j a^k .
  \label{10}
\end{equation}
It is useful to express the three $(1,1)$ moduli in terms of $V$ and
two additional independent moduli. This can be accomplished by
defining the scaled shape moduli
\begin{equation}
  b^i=V^{-1/3}a^i , \quad i=1,2,3 \ .
  \label{11}
\end{equation}
It follows from \eqref{10} that they satisfy the constraint
\begin{equation}
d_{ijk}b^ib^jb^k=6
\label{12}
\end{equation}
and, hence, represent only two degrees of freedom. Finally, note that
all moduli defined thus far, that is, the $a^i$, $V$ and $b^i$ are
functions of the five coordinates $x^{\alpha}$, $\alpha=0,\dots,3,11$
of $M_4 \times S^1/\Z_2$, where $x^{11}\in[0,\pi \rho]$.

\subsubsection{The Observable Sector Gauge Bundle} 

On the observable orbifold plane, the vector bundle $\Vvis$ on $X$
is chosen to be holomorphic with structure group $SU(4)\subset E_8$,
thus breaking
\begin{equation}
  E_8 \longrightarrow Spin(10) .
  \label{13}
\end{equation}
To preserve $N=1$ supersymmmetry in four-dimensions, $\Vvis$ must be
both slope-stable and have vanishing slope~\cite{Braun:2005zv,Braun:2006ae}. In the context of
this paper, these constraints are most easily examined in the $d=4$
effective theory and, hence, will be discussed in Section 3 below.  Finally, when
two \emph{flat} Wilson lines are turned on, each generating a
different $\Z_3$ factor of the $\Z_3 \times \Z_3$ holonomy of $X$, the
observable gauge group can be further broken to \footnote{As discussed in \cite{Barger:2008wn,FileviezPerez:2009gr}, the two U(1) factor groups depend on the explicit choice of Wilson lines. For the renormalization group analysis of the low-energy d=4 theory, it is more convenient to choose $U(1)_{T_{3R}}  \times U(1)_{B-L}$. However, since this is not our concern in this paper, we present the more canonical choice $U(1)_{Y}  \times U(1)_{B-L}$.} 
\begin{equation}
  Spin(10) 
  \longrightarrow 
  SU(3)_C \times SU(2)_L \times U(1)_Y \times U(1)_{B-L} .
  \label{17}
\end{equation}

Our analysis will require the Chern classes of $\Vvis$. Since the
structure group is $SU(4)$, it follows immediately that
$\rank(\Vvis)=4$ and $c_1(\Vvis)=0$. The heterotic standard model
is constructed so as to have the observed three chiral families of
quarks/leptons and, hence, $\Vvis$ is constructed so that
$c_3(\Vvis)=3$. Finally, we found in~\cite{Braun:2005nv,
  Braun:2005zv} that
\begin{equation}
  c_2(\Vvis) = 
  \frac{1}{v^{2/3}} \big(
  \omega_1 \wedge \omega_1+4~\omega_2 \wedge \omega_2
  + 4~\omega_1 \wedge \omega_2 
  \big).
\label{14}
\end{equation}
Here, and below, it will be useful to note the following. Let
${\cal{V}}$ be an arbitrary vector bundle on $X$ with structure group ${\cal{G}}$,
and ${\cal{F}}^{\cal{V}}$ the associated Lie algebra valued two-form
gauge field strength. If the generators of ${\cal{G}}$ are chosen to be
hermitian, then
\begin{equation}
  \frac{1}{8\pi^{2}}
  \tr_{\cal{G}} {\cal{F}^{\cal{V}}} \wedge {\cal{F}}^{\cal{V}}
  =ch_{2}({\cal{V}})
  =\frac{1}{2}c_{1}({\cal{V}}) 
  \wedge c_{1}({\cal{V}})-c_{2}({\cal{V}}) ,
\label{15}
\end{equation}
where $ch_{2}({\cal{V}})$ is the second Chern character of
${\cal{V}}$. Furthermore, we denote by $\tr_{\cal{G}}$ the trace in the fundamental
representation of the structure group ${\cal{G}}$ of the bundle. When applied
to the vector bundle $V^{(1)}$ in the observable sector, it follows
from $c_{1}(V^{(1)})=0$ that
\begin{equation}
  c_{2}(V^{(1)})
  =
  -\frac{1}{8\pi^{2}}\tr_{SU(4)} F^{(1)} \wedge F^{(1)} 
  =
  -\frac{1}{16\pi^{2}}\tr_{E_8} F^{(1)} \wedge F^{(1)} 
  ,
  \label{16}
\end{equation}
where $F^{(1)}$ is the gauge field strength for the visible sector
bundle $V^{(1)}$ and $\tr_{E_{8}}$ indicates the trace is over the fundamental $\Rep{248}$ representation of $E_{8}$. Note that the conventional normalization of the
trace $\tr_{E_8}$ includes a factor of
$\tfrac{1}{30}$, the inverse of the dual Coxeter number of $E_8$. We have expressed $c_{2}(V^{(1)})$ in terms of $\tr_{E_{8}}$ since the fundamental $SU(4)$ representation must be embedded into the adjoint representation of $E_{8}$ in the observable sector. 

For the visible sector bundle $V^{(1)}$ with structure group $SU(4)$, the group-theoretic embedding is
simply the standard $SU(4)\subset SU(9)\hookrightarrow E_8$.

\subsubsection{The Hidden Sector Gauge Bundle}

On the hidden orbifold plane, we will consider more general vector bundles and group
embeddings. Specifically, in this paper, we will restrict any choice of  hidden sector bundle to have the generic form of a Whitney sum 
\begin{equation}
V^{(2)}={\cal{V}}_{N} \oplus {\cal{L}}~, \qquad {\cal{L}}=\bigoplus_{r=1}^R L_r 
\label{dude1}
\end{equation}
where ${\cal{V}}_{N}$ is a slope-stable, non-Abelian bundle and 
each $L_{r}$, $r=1,\dots,R$ is a holomorphic line bundle with  structure group $U(1)$. Note that a subset of hidden sector vector bundles might have no non-Abelian factor at all, being composed entirely of the sum of one or more line bundles. On the other hand, one could choose the hidden sector bundle to be composed entirely of a non-Abelian vector bundle, that is, with no line bundle factors. Should the hidden sector bundle contain a non-Abelian factor, one could generically choose it to possess an arbitrary structure group. However, in this paper, for specificity, we will assume that the structure group of the non-Abelian factor is $SU(N)$ for some $N$. The explicit embeddings of the $SU(N)$ and individual $U(1)$ structure groups into the hidden sector $E_{8}$ gauge group will be discussed below. Finally, to preserve $N=1$ supersymmmetry in four-dimensions, $V^{(2)}$, being a Whitney sum of vector bundles,  must be poly-stable--generically with vanishing slope (but, importantly, see Section 3.2.1 below). As with the observable sector vector bundle, these constraints are most easily examined in the $d=4$ effective theory and, hence, will be discussed in Section 3 below. Let us first examine the non-Abelian factor.\\

\noindent $\bullet$ {\bf Hidden Sector $SU(N)$ Factor}\\

Since the structure group of ${\cal{V}}_{N}$ is $SU(N)$, it follows immediately that ${\rm rank}({\cal{V}}_{N})=N$ and $c_{1}({\cal{V}}_{N})=0$. The precise form of the second Chern class depends on the type of ${\cal{V}}_{N}$ bundle one chooses. Since this bundle is no longer constrained to give any particular spectrum it, and its associated second Chern class, can be quite general. 
The generic form for the second Chern class is given by
\begin{equation}
c_{2}({\cal{V}}_{N})=\frac{1}{v^{2/3}}(c_{N}^{ij}\omega_{i} \wedge \omega_{j})
\label{tase1}
\end{equation}
where $c_{N}^{ij}$ are, a priori, arbitrary real coefficients. Finally, note from \eqref{15} that since $c_{1}({\cal{V}}_{N})=0$,
\begin{equation}
ch_{2}({\cal{V}}_{N})=-c_{2}({\cal{V}}_{N}) \ .
\label{tase2}
\end{equation}
Let us now consider the line bundle factors.\\

\noindent $\bullet$ {\bf Hidden Sector Line Bundles}\\

Let us briefly review the properties of holomorphic line bundles on
our specific geometry. Line bundles are classified by the divisors of
$X$ and, hence, equivalently by the elements of the integral
cohomology
\begin{equation}
  H^2(X,\Z)= 
  \left\{a \omega_1+b \omega_2+c \omega_3 
    |a,b,c \in \Z 
  \right\} .
\label{18}
\end{equation}
It is conventional to denote the line bundle associated with the
element $a \omega_1+b \omega_2+c \omega_3$ of
$H^2(X,\Z)$ as
\begin{equation}
  \Ocal_X(a,b,c) .
  \label{19}
\end{equation}
Furthermore, in order for these bundles to arise from $\Z_3 \times
\Z_3$ equivariant line bundles on the covering space of $X$, they must
satisfy the additional constraint that
\begin{equation}
  a+b = 0 \tmod 3 .
  \label{20}
\end{equation}
Finally, as discussed in~\cite{Braun:2013wr}, for the purposes of constructing a heterotic gauge bundle from $\Ocal_X(a,b,c)$, \eqref{20} is the only constraint required on the integers $a,b,c$. Specifically, it is not necessary to impose that these integers be even for there to exist a spin structure on $V^{(2)}$.

We will choose the Abelian factor of the hidden bundle to be
\begin{equation}
  {\cal{L}} = \bigoplus_{r=1}^R L_r
  ,\quad 
  L_r=\Ocal_X(\ell^1_r, \ell^2_r, \ell^3_r)
  \label{21}
\end{equation}
where
\begin{equation}
  (\ell^1_r+\ell^2_r) \tmod 3 = 0
  , \quad r=1,\dots,R 
  \label{22}
\end{equation}
for any positive integer $R$. The structure group is $U(1)^R$, where
each $U(1)$ factor has a specific embedding into the hidden sector
$E_8$ gauge group. It follows from the definition that
$\rank({\cal{L}})=R$ and that the first Chern class is
\begin{equation}
  c_1({\cal{L}})
  =\sum_{r=1}^{R}c_1(L_r), \quad c_{1}(L_{r})=
  \frac{1}{v^{1/3}}  (\ell^1_r \omega_1 + \ell^2_r \omega_2 + \ell^3_r \omega_3) 
  .
\label{23}
\end{equation}
Note that since ${\cal{L}}$ is a sum of holomorphic line bundles,
$c_2({\cal{L}})=c_3({\cal{L}})=0$. However, the relevant quantity for the
hidden sector vacuum is the second Chern character defined in
\eqref{15}. For ${\cal{L}}$ this becomes
\begin{equation}
  ch_2({\cal{L}})
  = \sum_{r=1}^R ch_2(L_r)
\label{24}
\end{equation}
Since $c_2(L_r)=0$, it follows that
\begin{equation}
  ch_2(L_r)=2a_r\frac{1}{2}c_1(L_r) \wedge c_1(L_r) 
  \label{25}
\end{equation}
where
\begin{equation}
  a_r=\frac{1}{4\cdot 30} \tr_{E_8} Q_r^2
  \label{26}
\end{equation}
with $Q_r$ the generator of the $r$-th $U(1)$ factor embedded into
the $\Rep{248}$ representation of the hidden sector $E_8$. Computation of this $a$ coefficient depends on the choice of the hidden sector and will be discussed in more detail in Section 3.3.\\

The relevant topological object in the analysis of this paper will be the second Chern character of the complete hidden sector bundle
\begin{equation}
ch_{2}(V^{(2)})=ch_{2}({\cal{V}}_{N} \oplus {\cal{L}})=ch_{2}({\cal{V}}_{N})+ch_{2}({\cal{L}})  \ .
\label{dude3}  
\end{equation}
Using \eqref{tase2} and \eqref{24},\eqref{25} this becomes
\begin{equation}
ch_{2}(V^{(2)})= -c_{2}({\cal{V}}_{N})+ \sum_{r=1}^R a_r c_1(L_r) \wedge c_1(L_r)  
\label{dude4}
\end{equation}
with $a_{r}$ given in \eqref{26}. Note from \eqref{15}, the explicit embedding of the structure group of $V^{(2)}$ into $E_{8}$ and \eqref{dude4} that
\begin{equation}
\frac{1}{16 \pi^2} \tr_{E_8} F^{(2)} \wedge F^{(2)}=-c_{2}({\cal{V}}_{N})+ \sum_{r=1}^R a_r c_1(L_r) \wedge c_1(L_r)  \ .
\label{dude5}
\end{equation}

\subsubsection{Bulk Space Five-Branes}

In addition to the holomorphic vector bundles on the observable and
hidden orbifold planes, the bulk space between these planes can
contain five-branes wrapped on two-cycles ${\cal{C}}_2^{(n)}$,
$n=1,\dots,N$ in $X$. Cohomologically, each such five-brane is
described by the $(2,2)$-form Poincare dual to ${\cal C}_2^{(n)}$,
which we denote by $W^{(n)}$. Note that to preserves $N=1$
supersymmetry in the four-dimensional theory, these curves must be
holomorphic and, hence, each $W^{(n)}$ is an effective class.

\subsubsection{Anomaly Cancellation} 

As discussed in~\cite{Lukas:1998tt,Donagi:1998xe}, anomaly cancellation in heterotic M-theory requires that
\begin{equation}
  \sum_{n=0}^{N+1}J^{(n)}=0 ,
  \label{27}
\end{equation}
where 
\begin{equation}
  \label{28}
  \begin{split}
    J^{(0)}=&\;
    -\frac{1}{16 \pi^2}
    \Big( \tr_{E_8} F^{(1)} \wedge F^{(1)}
    -\frac{1}{2}\tr_{SO(6)} R \wedge R \Big) \\[1ex]
    J^{(n)}=&\;
    W^{(n)}, \quad n=1,\dots,N, \\[1ex]
    J^{(N+1)}=&\;
    -\frac{1}{16 \pi^2}
    \Big( \tr_{E_8} F^{(2)} \wedge F^{(2)}
    -\frac{1}{2}\tr_{SO(6)} R \wedge R \Big) \\
  \end{split}
\end{equation}
Note that the indices $n=0$ and $n=N+1$ denote the observable and hidden sector domain walls respectively, and {\it not} the location of a five-brane.
Using \eqref{6}, \eqref{16} and
\eqref{dude5}, the anomaly cancellation condition can be
expressed as
\begin{equation}
  c_2(TX)-c_2(\Vvis)-c_{2}({\cal{V}}_{N})
  +\sum_{r=1}^R  a_r c_1(L_r) \wedge c_1(L_r) - W 
  = 0 ,
  \label{29}
\end{equation}
where $W=\sum_{n=1}^N W^{(n)}$ is the total five-brane class.

Condition \eqref{29} is expressed in terms of four-forms in
$H^{4}(X,\R)$. We find it easier to analyze its consequences by
writing it in the dual homology space $H_2(X,\R)$. In this case, the
coefficient of the $i$-th vector in the basis dual to
$(\omega_1,\omega_2,\omega_3)$ is given by wedging each term in
\eqref{29} with $\omega_i$ and integrating over $X$. Using \eqref{5},
\eqref{14} and the intersection numbers
\eqref{3}, \eqref{4} gives
\begin{equation}
  \frac{1}{v^{1/3}}
  \int_X
  \big( c_2(TX)-c_2(\Vvis) \big)
  \wedge \omega_i=
  \big( \tfrac{4}{3},\tfrac{7}{3},-4 \big)_i
  ,\quad 
  i = 1,2,3.
\label{30}
\end{equation}
For $c_{2}({\cal{V}}_{N})$, it follows from \eqref{3} and \eqref{tase1} that
\begin{equation}
  \frac{1}{v^{1/3}}
  \int_X
  \big( -c_2({\cal{V}}_{N}) \big)
  \wedge \omega_i=-d_{ijk}c_{N}^{jk} \ .
\label{tase3}
\end{equation}
Similarly, \eqref{3},\eqref{4} and \eqref{23} imply
\begin{equation}
 \frac{1}{v^{1/3}}  \int_X{c_1(L_r) \wedge c_1(L_r)\wedge \omega_i}= 
  d_{ijk} \ell^j_r \ell^k_r \ , \quad i=1,2,3 .
  \label{31}
\end{equation}
Defining
\begin{equation}
  W_i = \frac{1}{v^{1/3}} \int_X W \wedge \omega_i \ ,
  \label{32}
\end{equation}
it follows that the anomaly condition \eqref{29} can be expressed as 
\begin{equation}
  W_i= \big( \tfrac{4}{3},\tfrac{7}{3},-4\big)\big|_i-d_{ijk}c_{N}^{jk}
  +\sum_{r=1}^R a_r d_{ijk} \ell^j_r \ell^k_r \geq 0 \ , \quad i=1,2,3  .
\label{33}
\end{equation}
The positivity constraint on $W$ follows from the requirement that it
be an effective class to preserve $N=1$ supersymmetry.

Finally, it is useful to define the charges
\begin{equation}
  \beta^{(n)}_i = 
  \frac{1}{v^{1/3}}
  \int_X J^{(n)} \wedge \omega_i \ , \quad i=1,2,3  .
\label{34}
\end{equation}
For example, it follows from \eqref{28}, using \eqref{5}, \eqref{6},
\eqref{14}, \eqref{16} and the intersection numbers \eqref{3},
\eqref{4}, that
\begin{equation}
  \beta^{(0)}_i = 
  \big( \tfrac{2}{3},-\tfrac{1}{3},4 \big)\big|_i \ .
  \label{35}
\end{equation}
Note that the anomaly condition \eqref{27} can now be expressed as
\begin{equation}
  \sum_{n=0}^{N+1}\beta^{(n)}_i=0 \ .
  \label{36}
\end{equation}

\subsection{The d=5$\rightarrow$d=4 Compactification}
\label{sec:d5to4}

\subsubsection{The Linearized Double Domain Wall}

The five-dimensional effective theory of heterotic M-theory, obtained
by dimensionally reducing Horava-Witten theory on the above Calabi-Yau
threefold, admits a BPS double domain wall solution with five-branes
in the bulk space \cite{Lukas:1998yy,Donagi:1999gc,Lukas:1998tt,Lukas:1998hk,Lukas:1999kt,Lukas:1997fg}. This solution depends on the previously defined
moduli $V$ and $b^i$ as well as the $a$, $b$ functions of the
five-dimensional metric
\begin{equation}
  ds_5^2=a^2dx^{\mu}dx^{\nu}\eta_{\mu\nu}+b^2dy^2 \ ,
  \label{37}
\end{equation}
all of which are dependent on the five coordinates $x^{\alpha}$,
$\alpha=0,\dots,3,11$ of $M_4 \times S^1/\Z_2$. Denoting the reference radius of $S^{1}$ by $\rho$, then $x^{11} \in [0,\pi \rho]$. These moduli can all
be expressed in terms of functions $f^i,~i=1,2,3$ satisfying the equations
\begin{equation}
  d_{ijk}f^jf^k=H_i \ , 
  \label{38}
\end{equation}
where each $H_i$ is a linear function of $z=\frac{x^{11}}{\pi\rho}$
with $z \in [0,1]$ and whose exact form depends on the number and
position of five-branes in the bulk space. As a simple, and relevant,
example, let us consider the case when there are no five-branes in the
vacuum. Then
\begin{equation}
  H_i=4k\epsilon_S'\beta^{(0)}_i
  \left( z-\frac{1}{2}\right) + k_i ,
  \label{39}
\end{equation}
where 
\begin{equation}
  \epsilon'_S = \pi \epsilon_{S}~ \ , ~
  \epsilon_{S}= \left(\frac{\kappa_{11}}{4\pi} \right)^{2/3}\frac{2\pi\rho}{v^{2/3}} 
  \label{40}
\end{equation}
and the charge $\beta^{(0)}_i$ is given in \eqref{35}. The $k$, $k_i$ are independent of $z$, but otherwise 
arbitrary functions of the four-dimensional moduli.

We are unable to give an exact analytic solution of \eqref{38} and
\eqref{39}. However, one can obtain an approximate solution by
expanding to linear order in $\epsilon_S'\beta^{(0)}_i\big(
z-\frac{1}{2}\big)$. It is clear from \eqref{39} that this
approximation will be valid under the conditions that
\begin{equation}
2\epsilon_S'|k\beta^{(0)}_i| \ll |k_i|
\label{41}
\end{equation}
for each $i=1,2,3$. This \emph{linearized} solution was discussed in
detail in~\cite{Lukas:1998hk,Lukas:1999kt,Lukas:1997fg}. Here we present only the results required in this
paper.  For an arbitrary dimensionless function $f$ of the five $M_4
\times S^1/\Z_2$ coordinates, define its average over the $S^1/\Z_2$
orbifold interval as
\begin{equation}
\langle f \rangle_{11}=\frac{1}{\pi \rho}\int_0^{\pi\rho}{dx^{11}f} \ ,
\label{42}
\end{equation}
where $\rho$ is a reference length. Then $\langle f \rangle_{11}$ is
a function of the four coordinates $x^{\mu}$, $\mu=0,\dots,3$ of $M_4$
only. The linearized solution is expressed in terms of orbifold
average functions
\begin{equation}
  V_0=\langle V \rangle_{11} , \quad 
  b^i_0=\langle b^i \rangle_{11} , \quad 
  \frac{\Rhat_0}{2}=\langle b \rangle_{11} .
\label{42a}
\end{equation}
We have defined $\frac{\Rhat_0}{2}=\langle b \rangle_{11}$ to
conform to specific normalization later in the paper. One then finds that the linearized solution specifies that
\begin{equation}
k=\frac{{\hat{R}}_{0}}{2V_{0}^{2/3}}~ , ~k_{i}=d_{ijk}b_{0}^{j}b_{0}^{k}V_{0}^{1/3} \ .
\label{NEW1}
\end{equation}

In terms of
these averaged moduli, the conditions \eqref{41} for the validity of
the linearized approximation can be written as
\begin{equation}
  \epsilon_S' \frac{\Rhat_0}{V_0}
  \left|\beta_i^{(0)}\right| \ll 
  d_{ijk}b_0^jb_0^k  \ ,
  \label{42b}
\end{equation}
where we have removed the absolute value of the right-hand side since all elements of $d_{ijk}$ given in \eqref{4} and each field $b^{i}$ are non-negative.
Equivalently, one can write
\begin{equation}
  \epsilon_S' \frac{\Rhat_0}{V_0} \ll 
  d_{ijk} b_0^j b_0^k \left|\beta^{(0)}_i\right|^{-1}
  \label{42B}
\end{equation}
for each $i=1,2,3$. These conditions are actually much simpler than they first appear. This can be seen by writing out each equation for$ i=1,2,3$ respectively. Using \eqref{35}, we find that
\begin{align}
i=1: \qquad q\quad &\frac{2}{3}\epsilon_S' \frac{\Rhat_0}{V_0} \ll \frac{2}{3}b_{0}^{1}b_{0}^{2}+\frac{1}{3}(b_{0}^{2})^{2}+2b_{0}^{2}b_{0}^{3}  \label{one}\\
i=2: \qquad \qquad &\frac{1}{3} \epsilon_S' \frac{\Rhat_0}{V_0} \ll \frac{1}{3}(b_{0}^{1})^{2}+\frac{2}{3}b_{0}^{1}b_{0}^{2}+2b_{1}^{2}b_{3}^{3} \label{two}\\
i=3: \qquad \qquad &4\epsilon_S' \frac{\Rhat_0}{V_0} \ll 2b_{0}^{1}b_{0}^{2} \label{three}
\end{align}
Clearly, if the equation for $i=3$ is satisfied then equations $i=1,2$ are automatically satisfied as well. Hence, one can replace the constraint \eqref{42B} by the simpler requirement that
\begin{equation}
2\epsilon_S' \frac{\Rhat_0}{V_0} \ll b_{0}^{1}b_{0}^{2}
\label{four} \ .
\end{equation}
%
%
%
%
%

Assuming that these conditions are fulfilled, the linearized solution
for $V$, $b^i$, $a$ and $b$ can be determined in terms of the orbifold
average functions. For example, assuming there are no five-branes in
the bulk space, the linearized solution for $V$ is given by
\begin{equation}
  V = V_0\left( 1+\epsilon_S'\frac{\Rhat_0}{V_0} 
    b^i_0\beta^{(0)}_i\big(z-\tfrac{1}{2}\big) \right) \ .
  \label{45}
\end{equation}
The linearized expressions for $b^i$ and $a,b$ are similar expansions
in the moduli dependent quantity
$(\epsilon_S'\frac{\Rhat_0}{V_0})b^i_0\beta^{(0)}_i$. It follows
that another check on the validity of these expansions is that
\begin{equation}
  \frac{1}{2} 
  \epsilon_S' \frac{\Rhat_0}{V_0}
  \left| b^i_0\beta^{(0)}_i \right| \ll 1 \  .
  \label{42A}
\end{equation}
%
%
%
However, using \eqref{four} it is straight forward to show that if constraint \eqref{42b} is satisfied then \eqref{42A} will be fulfilled automatically. To see this, note, using \eqref{35} and constraint \eqref{four}, that
\begin{equation}
 \frac{1}{2} 
  \epsilon_S' \frac{\Rhat_0}{V_0}
  \left| b^i_0\beta^{(0)}_i \right|= \frac{1}{2}  \epsilon_S' \frac{\Rhat_0}{V_0}\left| \frac{2}{3}b_{0}^{1}-\frac{1}{3}b_{0}^{2}+4b_{0}^{3} \right| \ll \frac{b_{0}^{1}b_{0}^{2}}{4}\left| \frac{2}{3}b_{0}^{1}-\frac{1}{3}b_{0}^{2}+4b_{0}^{3} \right|  \ .
\label{five}
\end{equation}
That is,
\begin{equation}
 \frac{1}{2} 
  \epsilon_S' \frac{\Rhat_0}{V_0}
  \left| b^i_0\beta^{(0)}_i \right| \ll  \left| \frac{1}{6}(b_{0}^{1})^{2}b_{0}^{2}-\frac{1}{12}b_{0}^{1}(b_{0}^{2})^{2}+b_{0}^{1}b_{0}^{2}b_{0}^{3} \right| \ .
\label{six}
\end{equation}
However, it follows from \eqref{4} for $d_{ijk}$ that
\begin{equation}
\frac{1}{6}d_{ijk}b_{0}^{i}b_{0}^{j}b_{0}^{k}= \frac{1}{6}(b_{0}^{1})^{2}b_{0}^{2}+\frac{1}{6}b_{0}^{1}(b_{0}^{2})^{2}+b_{0}^{1}b_{0}^{2}b_{0}^{3}=\frac{1}{6}(b_{0}^{1})^{2}b_{0}^{2}-\frac{1}{12}b_{0}^{1}(b_{0}^{2})^{2}+b_{0}^{1}b_{0}^{2}b_{0}^{3}+\frac{1}{4}b_{0}^{1}(b_{0}^{2})^{2} \ .
\label{sixA}
\end{equation}
Therefore
\begin{equation}
\left| \frac{1}{6}(b_{0}^{1})^{2}b_{0}^{2}-\frac{1}{12}b_{0}^{1}(b_{0}^{2})^{2}+b_{0}^{1}b_{0}^{2}b_{0}^{3} \right|=\left| \frac{1}{6}d_{ijk}b_{0}^{i}b_{0}^{j}b_{0}^{k}-\frac{1}{4}b_{0}^{1}(b_{0}^{2})^{2} \right| = \left| 1- \frac{1}{4}b_{0}^{1}(b_{0}^{2})^{2} \right| \ ,
\label{seven}
\end{equation}
where we have used expression \eqref{12}. However, it is clear from \eqref{six} that 
\begin{equation}
0 < \frac{1}{4}b_{0}^{1}(b_{0}^{2})^{2} < \frac{3}{2}
\label{sevenplus}
\end{equation}
and therefore
\begin{equation}
\frac{1}{2} 
  \epsilon_S' \frac{\Rhat_0}{V_0}
  \left| b^i_0\beta^{(0)}_i \right| \ll \left|1-\frac{1}{4}b_{0}^{1}(b_{0}^{2})^{2}  \right| < 1\ .
\label{eight}
\end{equation}
Hence, condition \eqref{42A} is automatically satisfied if constraint \eqref{42b} is. 

Thus far, we have considered the case when there are no five-branes in
the bulk space. Including an arbitrary number of five-branes in the
linearized BPS solution is straightforward and was presented
in~\cite{Lukas:1998yy,Donagi:1999gc,Lukas:1998tt}. Here, it will suffice to generalize the above discussion
to the case of one five-brane located at $z_1 \in [0,1]$. The
conditions for the validity of the linear approximation then break
into two parts. Written in terms of the averaged moduli, these are
\begin{equation}
  2\epsilon_S'\frac{\Rhat_0}{V_0}
  \left|
    \beta_i^{(0)} \big(z-\tfrac{1}{2}\big)
    -\frac{1}{2}\beta_i^{(1)}(1-z_1)^2
  \right|
  \ll 
  \left| d_{ijk} b_0^j b_0^k \right|
  , \quad z \in [0,z_1]
\label{45B}
\end{equation}
and 
\begin{equation}
  2\epsilon_S'\frac{\Rhat_0}{V_0}
  \left|
    (\beta_i^{(0)}+\beta_i^{(1)})
    \big(z-\tfrac{1}{2}\big)
    -\frac{1}{2}\beta_i^{(1)}z_1^2
  \right| 
  \ll 
  \left| d_{ijk} b_0^j b_0^k \right|
  , \quad z \in [z_1,1] .
\label{45C}
\end{equation}
Assuming these conditions are satisfied, the linearized solution for
$V$, $b_i$ and $a$, $b$ can be determined in each region. For example,
the linearized solution for $V$ is given by
\begin{equation}
  V=V_0\Big(1+  \epsilon_S' \frac{\Rhat_0}{V_0}
 b^i_0\Big[\beta_i^{(0)}\big(z-\tfrac{1}{2}\big)-\frac{1}{2}\beta_i^{(1)}(z_1-1)^2
  \Big]\Big) , 
  \quad z\in[0,z_1]
\label{45D}
\end{equation}
and
\begin{equation}
  V=V_0\Big(1+  \epsilon_S' \frac{\Rhat_0}{V_0} 
b^i_0\Big[(\beta_i^{(0)}+\beta_i^{(1)})
  \big(z-\tfrac{1}{2}\big)
  -\frac{1}{2}\beta_i^{(1)}z_1^2
  \Big]\Big) , \quad z\in[z_1,1] \ .
\label{45b}
\end{equation}
It follows that the conditions for the validity of this linearized solution for $V$ are given by
\begin{equation}
 \epsilon_S' \frac{\Rhat_0}{V_0}
 b^i_0\Big|\beta_i^{(0)}\big(z-\tfrac{1}{2}\big)-\frac{1}{2}\beta_i^{(1)}(z_1-1)^2
  \Big| \ll 1, 
  \quad z\in[0,z_1]
\label{45DD}
\end{equation}
and
\begin{equation}
\epsilon_S' \frac{\Rhat_0}{V_0} 
b^i_0\Big|(\beta_i^{(0)}+\beta_i^{(1)})
  \big(z-\tfrac{1}{2}\big)
  -\frac{1}{2}\beta_i^{(1)}z_1^2
  \Big| \ll 1 , \quad z\in[z_1,1] \ .
\label{45bb}
\end{equation}
Note that if the five-brane is located near the hidden wall, that is,
$z_1\rightarrow 1$, conditions \eqref{45B} and \eqref{45C} for the
validity of the linear approximation both revert to \eqref{42b}, and, hence, inequality \eqref{four},  as
they must for consistency. Similarly, the conditions \eqref{45DD} and \eqref{45bb} for the validity of the solutions for $V$--as well as for the expansions of $b_{i}$ and $a$, $b$-- simply reduce to \eqref{42A}. Again, we find that conditions \eqref{45DD} and \eqref{45bb} will be automatically satisfied if the strong coupling constraints \eqref{45B} and \eqref{45C} are. For this reason, we henceforth consider the strong coupling constraints only.

When dimensionally reduced on this linearized BPS solution, the
four-dimensional functions $a_0^i$, $V_0$, $b_0^i$ and $\Rhat_0$
will become moduli of the $d=4$ effective heterotic M-theory. The
geometric role of $a_0^i$ and $V_0, b_0^i$ will remain the same as
above---now, however, for the averaged Calabi-Yau threefold. For
example, the dimensionful volume of the averaged Calabi-Yau manifold
will be given by $vV_0$. The new dimensionless quantity $\Rhat_0$
will be the length modulus of the orbifold. The dimensionful length of
$S^1/\Z_2$ is given by $\pi \rho \Rhat_0$. Finally, since the
remainder of this paper will be within the context of the $d=4$
effective theory, we will, for simplicity, drop the subscript ``$0$''
on all moduli henceforth.

\section{The d=4 $E_8 \times E_8$ Effective Theory}

When $d=5$ heterotic M-theory is dimensionally reduced to four
dimensions on the linearized BPS double domain wall with five-branes,
the result is an $N=1$ supersymmetric effective four-dimensional
theory with (potentially spontaneously broken) $E_8 \times E_8$ gauge group. The
Lagrangian will break into two distinct parts. The first contains
terms of order $\kappa_{11}^{2/3}$ in the eleven-dimensional Planck
constant $\kappa_{11}$, while the second consists of terms of order
$\kappa_{11}^{4/3}$.

\subsection{The $\kappa_{11}^{2/3}$ Lagrangian}

This Lagrangian is well-known and was presented in~\cite{Lukas:1997fg}. Here we
discuss only those properties required in this paper. In four
dimensions, the moduli must be organized into the lowest components of
chiral supermultiplets. Here, we need only consider the real part of
these components. Additionally, one specifies that these chiral
multiplets have \emph{canonical} K\"ahler potentials in the effective
Lagrangian. The dilaton is simply given by
\begin{equation}
  \re S=V \ .
  \label{46}
\end{equation}
However, neither $a^i$ nor $b^i$ have canonical kinetic energy. To
obtain this, one must define the rescaled moduli
\begin{equation}
  t^i = 
  \Rhat b^i = 
  \Rhat V^{-1/3} a^i \ ,
  \label{47}
\end{equation}
where we have used \eqref{11}, and choose the complex K\"ahler moduli
$T^i$ so that
\begin{equation}
  \re T^i = t^i \ .
\label{48}
\end{equation}
Denote the real modulus specifying the location of the $n$-th
five-brane in the bulk space by $z_n=\frac{x^{11}_n}{\pi \rho}$ where
$n=1,\dots,N$. As with the K\"ahler moduli, it is necessary to define
the fields
\begin{equation}
  \re Z^n = \beta_i^{(n)} t^i z_n \ .
\label{49}
\end{equation}
These rescaled $Z^n$ five-brane moduli have canonical kinetic
energy.

The gauge group of the $d=4$ theory has two $E_8$ factors, the first
associated with the observable sector and the second with the hidden
sector.   As discussed previously, both vector bundles must be chosen so as to preserve $N=1$ supersymmetry in four-dimensions. We now explicitly discuss the conditions under which this will be true. We begin with the observable sector.\\

\noindent $\bullet$ {\bf Stability of the Observable Sector Vector Bundle}\\

To preserve $N=1$ supersymmmetry in
four-dimensions the holomorphic $SU(4)$ vector bundle $\Vvis$
associated with the observable $E_8$ gauge group must be both
slope-stable and have vanishing slope~\cite{MR88i:58154,MR86h:58038,Green:1987mn}.  The slope of any
bundle or sub-bundle $\cal{F}$ is defined as
\begin{equation}
  \mu({\cal{F}})=
  \frac{1}{\rank({\cal{F}})v^{2/3}} 
  \int_X{c_1(\cal{F})\wedge \omega \wedge \omega} \ ,
  \label{50}
\end{equation}
where $\omega$ is the K\"ahler form in \eqref{8}---now, however,
written in terms of the $a^i$ moduli \emph{averaged} over
$S^1/\Z_2$. Since $c_1(\Vvis)=0$, $\Vvis$ has vanishing
slope. But, is it slope-stable? As proven in detail in~\cite{Braun:2006ae}, this
will be the case in a subspace of the K\"ahler cone defined by seven
inequalities required for all sub-bundles of $V^{(1)}$ to have negative
slope. These can be slightly simplified into the statement that the moduli $a^{i},~i=1,2,3$ must satisfy at least one of the two inequalities
\begin{figure}[htbp]
  \centering
  \includegraphics[width=0.9\textwidth]{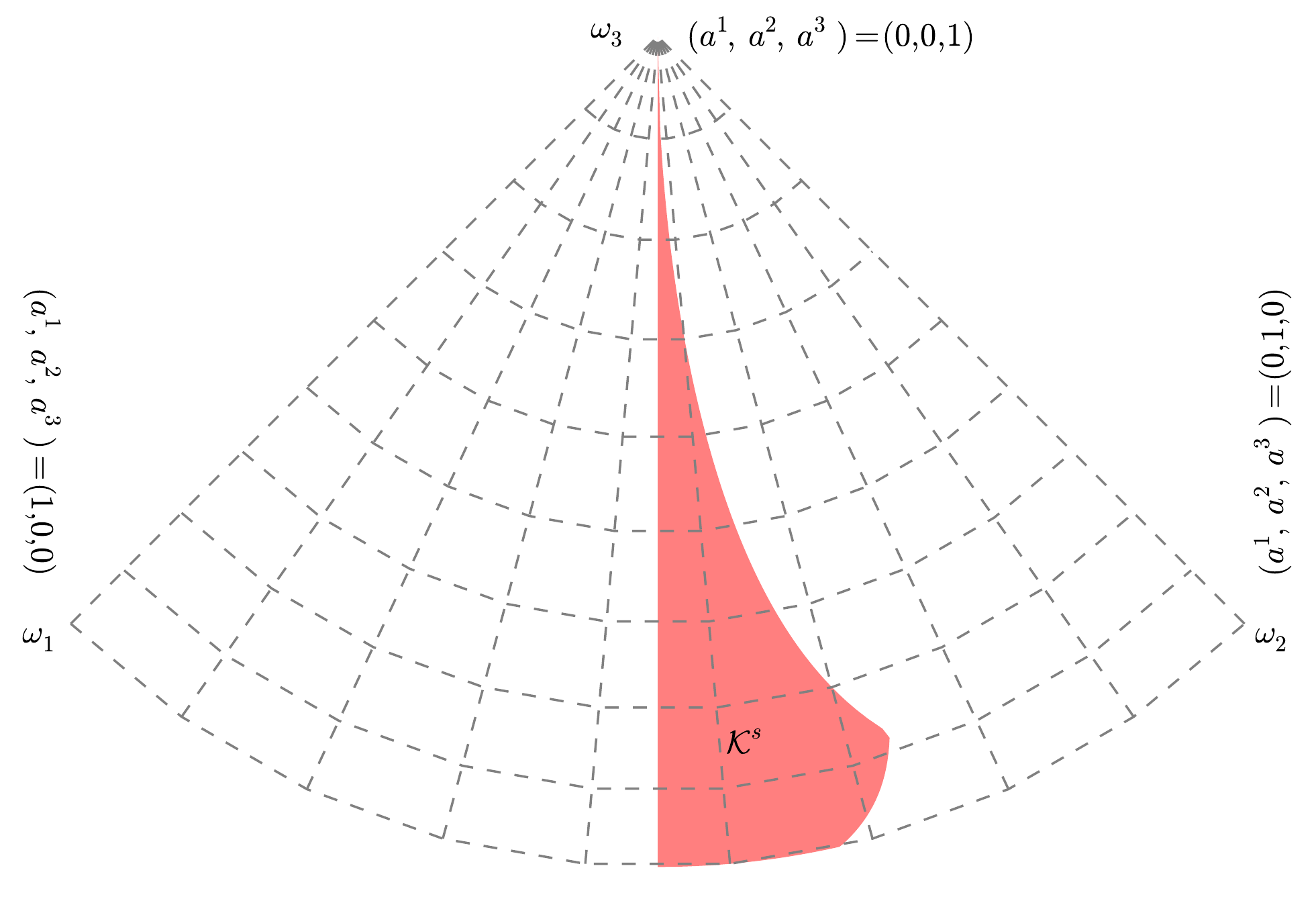}
  \caption{The visible sector stability region in the K\"ahler cone.}
  \label{fig:starmap}
\end{figure}
\begin{equation}\label{51}
  \begin{gathered}
    \left(
      a^1
      < 
      a^2
      \leq 
      \sqrt{\tfrac{5}{2}} a^1
      \quad\text{and}\quad
      a^3
      <
      \frac{
        -(a^1)^2-3 a^1 a^2+ (a^2)^2
      }{
        6 a^1-6 a^2
      } 
    \right)
    \quad\text{or}\\
    \left(
      \sqrt{\tfrac{5}{2}} a^1
      <
      a^2
      <
      2 a^1
      \quad\text{and}\quad
      \frac{
        2(a^2)^2-5 (a^1)^2
      }{
        30 a^1-12 a^2
      }
      <
      a^3
      <
      \frac{
        -(a^1)^2-3 a^1 a^2+ (a^2)^2
      }{
        6 a^1-6 a^2
      }
    \right) 
  \end{gathered}
\end{equation}

The subspace $\Kcone^s$ satisfying \eqref{51} is a full-dimensional
subcone of the K\"ahler cone $\Kcone$ defined in \eqref{7}. 
It is a cone because the inequalities are homogeneous. In other words, only the angular part of the K\"ahler moduli are constrained, but not the overall volume. 
Hence, it is best displayed as a two-dimensional ``star map'' as seen by an
observer at the origin. This is shown in \autoref{fig:starmap}. For
K\"ahler moduli restricted to this subcone, the four-dimensional low
energy theory in the observable sector is $N=1$ supersymmetric.\\

\noindent $\bullet$ {\bf Poly-Stability of the Hidden Sector Vector Bundle}\\

To preserve $N=1$ supersymmmetry in
four-dimensions, the hidden sector vector bundle must satisfy two conditions, First, since it is generically a Whitney sum, the vector bundle must be poly-stable. That is, each factor of the Whitney sum must be slope-stable and, in addition, all factors in the sum must have the same slope. Second, generically, this slope must vanish identically--but with one important caveat discussed in Section 3.3.2. In order to make this more concrete, we now present three non-trivial examples to illustrate the these two conditions. As a first example, let us choose 
\begin{enumerate}

\item $V^{(2)}={\cal{V}}_{N}$:\\
First, in this case, since for a single vector bundle slope-stability implies poly-stability, one need only check that ${\cal{V}}_{N}$ is slope-stable. For example, one could choose ${\cal{V}}_{N}$ to be identical to the $SU(4)$ bundle in the observable sector, $V^{(1)}$, presented above. Note that, since we are restricting all hidden sector non-Abelian bundles to have structure group $SU(N)$, it follows that $\mu({\cal{V}}_{N})$ must vanish, thus satisfying the second condition for $N=1$ supersymmetry. As with the observable sector bundle $SU(4)$ bundle, stability of a generic non-Abelian vector bundle will only occur within a specific region of K\"ahler moduli space.

\item $V^{(2)}=L$:\\
As in the previous case, one need only check that the line bundle $L$ is slope-stable, which will imply poly-stability. Fortunately, every line bundle is trivially slope-stable, so any line bundle can be used. It is important to note that the slope of a line bundle which appears as a lone factor in the Whitney sum has, a priori, no further constraints--that is, $\mu(L)$ need not vanish. Using \eqref{50}, \eqref{23} and \eqref{4}, it follows that the slope of an arbitrary line bundle specified by $L = \Ocal_X(\ell^1,\ell^2,\ell^3)$ is given by
\begin{multline}
\mu(L)=d_{ijk}\ell^{i}a^{j}a^{k}
\\
=\frac{1}{3}\Big( a^{2}(a^{2}+6a^{3})\ell^{1}+(a^{1})^{2}\ell^{2}+6a^{1}a^{3}\ell^{2} +2a^{1}a^{2}(\ell^{1}+\ell^{2}+3\ell^{3}) \Big).
\label{otter1}
\end{multline}
That is, its value is a highly specific function of the K\"ahler moduli. We will discuss the requirements that such a line bundle lead to four-dimensional $N=1$ supersymmetry in Section 3.2.2 below.

\item $V^{(2)}={\cal{V}}_{N} \oplus L:$\\
As specified above, the non-Abelian vector bundle ${\cal{V}}_{N}$ must be slope-stable in a region of K\"ahler moduli space. Furthermore, since we are restricting the structure group in our discussion to be $SU(N)$, it follows that $\mu({\cal{V}}_{N})=0$. As we just indicated, any line bundle $L$ will be slope-stable everywhere in K\"ahler moduli space. 
However, the full Whitney sum $V^{(2)}={\cal{V}}_{N} \oplus L$ will be poly-stable--and, hence, preserve $N=1$ supersymmetry--if and only if $\mu(L)=\mu({\cal{V}}_{N})=0$. That is, because of the existence of a non-Abelian $SU(N)$ factor, the line bundle $L$ now has the additional constraint that its slope vanish identically. It is clear from \eqref{otter1} that this will be the case only in a restricted region of K\"ahler moduli space. It follows that the full Whitney sum $V^{(2)}={\cal{V}}_{N} \oplus L$ will only be a viable hidden sector bundle if the region of stability of ${\cal{V}}_{N}$ has a non-vanishing intersection with the region where the slope of $L$ vanishes. This is a very non-trivial requirement.
To give a concrete example, let us choose  ${\cal{V}}_{N}=V^{(1)}$, where $V^{(1)}$ is the $SU(4)$ observable sector bundle specified above. Recall that the region of slope-stability of this bundle in K\"ahler moduli space is delineated by the inequalities in \eqref{51} and shown in Figure 1. Plotted in 3-dimensions, this region of slope-stability over a limited region of K\"ahler moduli space is shown in Figure 2(a). Furthermore, let us specify, for example, that $L = \Ocal_X(1,2,-3)$. Note that $L$ satisfies condition \eqref{20}, as it must. It follows from \eqref{otter1} that the region of moduli space in which $\mu(L)=0$ is given by the equation
\begin{equation}
\frac{2}{3}(a^{1})^{2}-4a^{1}a^{2}+\frac{1}{3}(a^{2})^{2}+4a^{1}a^{3}+2a^{2}a^{3}=0 \ .
\label{otter2}
\end{equation}
Plotted over a limited region of K\"ahler moduli space in 3-dimensions, the region where $\mu(L)=0$  is shown in Figure 2(b). Figure 2(c) then shows that these two regions have a substantial overlap in K\"ahler moduli space. Furthermore, since ${\cal{V}}_{N}$ was chosen to be $V^{(1)}$, it follows that Figure 2(c) also represents the overlap with the stability region of the observable sector vector bundle. We conclude that the specific choice of $V^{(2)}=V^{(1)} \oplus \Ocal_X(1,2,-3)$ is, potentially, a suitable choice for a poly-stable hidden sector vector bundle.

\begin{figure}[h!]
\begin{center}
\includegraphics[scale=0.34]{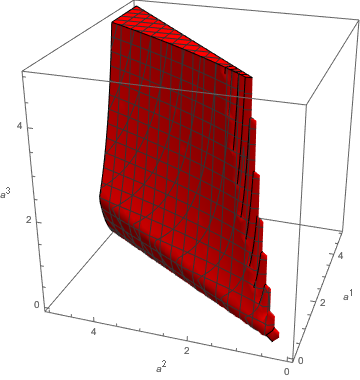}
\raisebox{1cm}{(a)}
\includegraphics[scale=0.34]{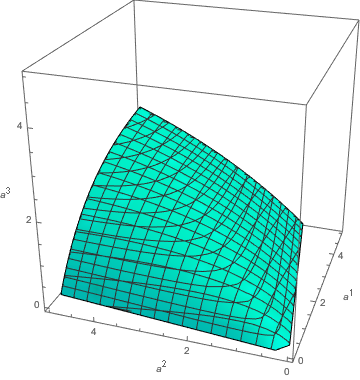}
 \raisebox{1cm}{(b)}
\includegraphics[scale=0.34]{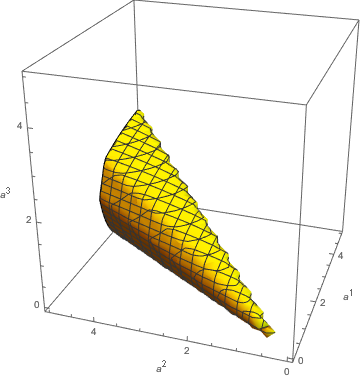}
\raisebox{1cm}{(c)}
\caption{The region of poly-stability for the hidden sector vector bundle $V^{(2)}=V^{(1)} \oplus \Ocal_X(1,2,-3)$. The red volume in Figure 2(a) is a sub-region of K\"ahler moduli space where the bundle $V^{(1)}$ is slope stable, whereas the green volume of Figure 2(b) is the sub-region where $\mu(\Ocal_X(1,2,-3))=0$. They have a substantial region of overlap in 
K\"ahler moduli space, indicated by the yellow volume in Figure 3(c).}
\label{default}
\end{center}
\end{figure}

\end{enumerate}
These three examples give the rules for constructing specific poly-stable vector bundles. They can easily be generalized to construct generic poly-stable Whitney sum hidden sector vector bundles.

\subsection{The $\kappa_{11}^{4/3}$ Lagrangian}

The terms in the BPS double domain wall solution proportional to
$\epsilon_S'$ lead to order $\kappa_{11}^{4/3}$ additions to the $d=4$
Lagrangian. These have several effects. The simplest is that the
five-brane location moduli now contribute to the definition of the
dilaton, which becomes
\begin{equation}
  \re S = V + \epsilon_S'\sum_{n=1}^{N}\beta_i^{(n)}t^iz_n^2 \ ,
  \label{52}
\end{equation}
where the fields $t^{i}$ are defined in \eqref{47}.
More profoundly, these $\kappa_{11}^{4/3}$ terms lead, first,  to threshold
corrections to the gauge coupling parameters and, second, to additions
to the Fayet-Iliopoulos (FI) term associated with any anomalous $U(1)$
factor in the low energy gauge group. Let us analyze these in
turn.

\subsubsection{Gauge Threshold Corrections}

The gauge couplings of the non-anomalous components of the $d=4$ gauge
group, in both the observable and hidden sectors, have been computed
to order $\kappa_{11}^{4/3}$ in~\cite{Lukas:1998hk}. Written in terms of the fields $b^{i}$ defined in \eqref{11} and including five-branes in the bulk
space, these are given by
\begin{equation}
  \frac{4\pi}{(g^{(1)})^2} \propto
  V (1+\epsilon_S' \frac{\Rhat}{2V} 
  \sum_{n=0}^{N}(1-z_n)^2 b^i \beta^{(n)}_i )
\label{62}
\end{equation}
and 
\begin{equation}
  \frac{4\pi}{(g^{(2)})^2} \propto
  V (1+\epsilon_S' \frac{\Rhat}{2V} 
  \sum_{n=1}^{N+1}z_n^2 b^i\beta^{(n)}_i)
  \label{63}
\end{equation}
respectively. The positive definite constant of proportionality is identical for both gauge couplings and is not relevant to the present discussion. It is important to note that the effective parameter of the $\kappa_{11}^{2/3}$ expansion in \eqref{62} and \eqref{63}, namely $\epsilon_S' \frac{\Rhat}{V}$, is identical to  1) the parameter appearing in \eqref{42b} (and its five-brane extension \eqref{45B} and \eqref{45C}) for the validity of the linearized approximation, as well as 2) the $\kappa_{11}^{2/3}$ expansion parameter for V presented in \eqref{45} (and its five-brane extension \eqref{45D} and \eqref{45DD}). That is, the effective strong coupling parameter of the $\kappa_{11}^{2/3}$ expansion is given by
\begin{equation}
\epsilon_{S}^{\rm eff}= \epsilon_S' \frac{\Rhat}{V} \ .
\label{pink1}
\end{equation}
We point out that this is, up to a constant factor of order one, precisely the strong coupling parameter presented in equation (1.3) of \cite{Banks:1996ss}.

Recall that $n=0$ and $n=N+1$ correspond to the observable and hidden sector domain walls--not to five-branes. Therefore, $z_{0}=0$ and $z_{N+1}=1$.
Using \eqref{28} and  \eqref{34}, one can evaluate the $\beta^{(n)}_i$ coefficients in terms of the the $a^{i}, i=1,2,3$ K\"ahler moduli defined in \eqref{8}. Rewritting the above expressions in terms of these moduli using  \eqref{6}, \eqref{8}, \eqref{10}, \eqref{11},
\eqref{24}, \eqref{25}, as well as redefining the five-brane moduli to be
\begin{equation}
  \lambda_n = 
  z_n-\frac{1}{2}
  , \quad 
  \lambda_n \in \left[-\tfrac{1}{2},\tfrac{1}{2}\right] ,
\label{pink2}
\end{equation}
we find that
\begin{equation}
  \label{64}
  \begin{split}
    \frac{4\pi}{(g^{(1)})^2} \propto \;&
    \frac{1}{6v}\int_X\omega \wedge \omega
    \wedge \omega 
    -\epsilon_S' \frac{\Rhat}{2V^{1/3}}
    \frac{1}{v^{1/3}}
    \\ &
    \times
    \int_X\omega \wedge 
    \left(
      -c_2(\Vvis) 
      +\frac{1}{2}c_2(TX)-\sum_{n=1}^{N}(\tfrac{1}{2}-\lambda_n)^2W^{(n)}  
    \right)
  \end{split}
\end{equation}
and 
\begin{equation}
  \label{65}
  \begin{split}
  \frac{4\pi}{(g^{(2)})^2} \propto \;&
  \frac{1}{6v}\int_X\omega \wedge \omega
  \wedge \omega 
  -\epsilon_S' \frac{\Rhat}{2V^{1/3}}
  \frac{1}{v^{1/3}}
  \\ &
  \times
  \int_X\omega \wedge 
  \left(
   -c_{2}({\cal{V}}_{N})+
    \sum_{r=1}^{R} a_{r} c_{1}(L_{r}) \wedge c_{1}(L_{r})
    +\frac{1}{2}c_2(TX)-\sum_{n=1}^{N}(\tfrac{1}{2}+\lambda_n)^2W^{(n)}  
    \right)
  \end{split}
\end{equation}
where $a_r$ is given in \eqref{26}. The first term on the right-hand
side, that is, the volume $V$ defined in \eqref{10}, is the order
$\kappa_{11}^{2/3}$ result. The remaining terms are the $\kappa_{11}^{4/3}$
M-theory corrections first presented in~\cite{Lukas:1998hk}.

Clearly, consistency of the $d=4$ effective theory requires both
$(g^{(1)})^2$ and $(g^{(2)})^2$ to be positive. It follows that the
moduli of the four-dimensional theory are constrained to satisfy
\begin{multline}
  \frac{1}{v}\int_X\omega \wedge \omega \wedge \omega -3\epsilon_S'
  \frac{\Rhat}{V^{1/3}} \frac{1}{v^{1/3}}\int_X\omega \wedge
  \big(-c_2(\Vvis) 
  \\
  +\frac{1}{2}c_2(TX)-\sum_{n=1}^{N}(\tfrac{1}{2}-\lambda_n)^2W^{(n)}  \big) > 0
  \label{66}
\end{multline}
and 
\begin{multline}
  \frac{1}{v}\int_X\omega \wedge \omega \wedge \omega -3\epsilon_S'
  \frac{\Rhat}{V^{1/3}} \frac{1}{v^{1/3}}\int_X\omega \wedge
  \big(-c_{2}({\cal{V}}_{N})+
    \sum_{r=1}^{R} a_{r} c_{1}(L_{r}) \wedge c_{1}(L_{r})
  \\
  +\frac{1}{2}c_2(TX)-\sum_{n=1}^{N}(\tfrac{1}{2}+\lambda_n)^2W^{(n)}  \big) > 0 .
  \label{67}
\end{multline}
%
One can use \eqref{3}, \eqref{4},
\eqref{5}, \eqref{8}, \eqref{14}, \eqref{tase1}, \eqref{23}  and \eqref{32} to rewrite
these expressions as
\begin{equation}
  \label{68}
  \begin{split}
    d_{ijk} a^i a^j a^k- 3 \epsilon_S' \frac{\Rhat}{V^{1/3}} \Big(
    -(\frac83 a^1 + \frac53 a^2 + 4 a^3)
    + \qquad& \\
    + 2(a^1+a^2) -\sum_{n=1}^{N}(\tfrac{1}{2}-\lambda_n)^2 a^i \;W^{(n)}_i
    \Big) &> 0 
  \end{split}
\end{equation}
and
\begin{equation}
  \label{69}
  \begin{split}
    d_{ijk} a^i a^j a^k- 3 \epsilon_S' \frac{\Rhat}{V^{1/3}}
    \Big(-d_{ijk}a^{i}c_{N}^{jk}+d_{ijk}a^i \sum_{r=1}^{R}a_r \ell^j_r \ell^k_r
    + \qquad& \\
    + 2(a^1+a^2) -\sum_{n=1}^{N}(\tfrac{1}{2}+\lambda_n)^2 a^i
    \;W^{(n)}_i \Big) &> 0
  \end{split}
\end{equation}
respectively.
It is of interest to compare the $(g^{(1)})^2$, $(g^{(2)})^2>0$
conditions calculated to order $\kappa_{11}^{4/3}$ in strongly coupled
heterotic M-theory, that is, \eqref{66} and \eqref{67}, to the
one-loop corrected conditions computed in the weakly coupled heterotic
string~\cite{Weigand:2006yj}. Assuming the same observable and hidden sector vector
bundles used in this paper, we find that the weakly coupled conditions for $(g^{(1)})^2$, $(g^{(2)})^2>0$, 
derived using equation (3.103) in~\cite{Weigand:2006yj}, are
identical to \eqref{66} and \eqref{67} if one replaces
\begin{equation}
g_{s}^{2}l_{s}^{4}~\longrightarrow~ \epsilon_{S}^{\prime}\frac{\hat{R}}{V^{1/3}}v^{2/3}
\label{red1}
\end{equation}
in the weak coupling formulas, where $g_{s}$ and $l_{s}=2 \pi \sqrt{\alpha^{\prime}}$ are the weak coupling parameter and the string length respectively and $ \epsilon_{S}^{\prime}$ is defined in \eqref{40}.

\subsubsection{Corrections to a Fayet-Iliopoulos Term}

In the heterotic standard model vacuum, the observable sector vector
bundle $\Vvis$ has structure group $SU(4)$. Hence, it does not lead
to an anomalous $U(1)$ gauge factor in the observable sector of the
low energy theory. However, the hidden sector bundle $\Vhid$ introduced above, in addition to a possible non-Abelian bundle ${\cal{V}}_{N}$, 
consists of a sum of line bundles with the additional structure group $U(1)^{R}$. Each $U(1)$ factor leads to
an anomalous $U(1)$ gauge group in the four-dimensional effective
field theory and, hence, an associated $D$-term.  Let $L_r$ be
any one of the irreducible line bundles of $\Vhid$.  The string one-loop corrected Fayet-Iliopoulos (FI) term for $L_{r}$ was
computed in~\cite{Blumenhagen:2005ga} within the context of the weakly coupled heterotic string. Comparing various results in the literature, it is straightforward to show that strong coupling results to order $\kappa_{11}^{4/3}$ can be obtained from string one-loop weak coupling expressions using the same replacement $g_{s}^{2}l_{s}^{4}~\longrightarrow~ \epsilon_{S}^{\prime}\frac{\hat{R}}{V^{1/3}}v^{2/3}$ presented in \eqref{red1}. Making this substitution, 
we find that the expression for the FI term associated with $L_{r}$ in strongly coupled heterotic M-theory  to order $\kappa_{11}^{4/3}$ is given by
\begin{equation}
 \label{54}
  FI_r =
  \frac{3}{16} \frac{ \epsilon_S \epsilon_r^2}{\kappa_{4}^{2}}
  \frac{1} {\Rhat V^{2/3}}
  \Big( \mu(L_r) + 
     \epsilon_S' \frac{\Rhat}{V^{1/3}} 
    \int_X c_1(L_r) \wedge 
   \big( J^{(N+1)}+\sum_{n=1}^{N} z_n^2 J^{(n)} \big) \Big) , 
\end{equation}
where $\mu(L_r)$ is given in \eqref{50}. We note that the $\kappa_{11}^{2/3}$ part of this expression is identical to that derived in \cite{Anderson:2009nt}.
Inserting \eqref{28}, \eqref{6}, \eqref{dude5} and, following the
conventions of~\cite{Blumenhagen:2005ga,Weigand:2006yj}, redefining the five-brane moduli as in \eqref{pink2},
we find that
\begin{multline}
  \label{56}
  FI_r = 
   \frac{3}{16} \frac{ \epsilon_S \epsilon_r^2}{\kappa_{4}^{2}}
  \frac{1} {\Rhat V^{2/3}}
  \Big( \mu(L_r) - 
  \epsilon_S' \frac{\Rhat}{V^{1/3}} 
  \\
  \int_X c_1(L_r)\wedge
  \big(-c_{2}({\cal{V}}_{N})+\sum_{s=1}^{R} a_{s} c_{1}(L_{s}) \wedge c_{1}(L_{s})
  +\frac{1}{2} c_2(TX) 
  -\sum_{n=1}^{N} (\tfrac{1}{2}+\lambda_n)^2 W^{(n)} \big) \Big) \ ,
\end{multline}
where $a_{s}$ is given in \eqref{26}. The first term on the right-hand
side, that is, the slope of $L_r$, is the
order $\kappa_{11}^{2/3}$ result. The remaining terms are the
$\kappa_{11}^{4/3}$ M-theory corrections first presented in~\cite{Lukas:1998hk}.
Note that the dimensionless parameter $ \epsilon_S'
\frac{\Rhat}{V^{1/3}}$ of the $\kappa_{11}^{4/3}$ term is identical to
the expansion coefficient of the linearized solution---when expressed
in term of the $a^i$ moduli---discussed in \autoref{sec:d5to4}. See,
for example, \eqref{42b}. Finally, recalling definition \eqref{50} of the slope, 
using \eqref{3}, \eqref{4}, \eqref{8},
\eqref{23}, \eqref{32} and noting from \eqref{5} that
\begin{equation}
 \frac{1}{v^{1/3}}  \int_X
  \frac{1}{2}c_2(TX) \wedge \omega_i=(2,2,0)_i 
  ,\quad i=1,2,3,
  \label{58}
\end{equation}
it follows that for each $L_r$ the associated Fayet-Iliopoulos factor $FI_{r}$ in  \eqref{56} can be written as
\begin{multline}
  \label{59} 
FI_{r}= \frac{3}{16} \frac{ \epsilon_S \epsilon_r^2}{\kappa_{4}^{2}}
  \frac{1} {\Rhat V^{2/3}} \Big(d_{ijk} \ell_r^i a^j a^k - 
  \epsilon_S' \frac{\Rhat}{V^{1/3}} 
  \\
 \big(-d_{ijk}\ell_{r}^{i}c_{N}^{jk}+d_{ijk}\ell_r^i\sum_{s=1}^{R}a_{s}\ell_{s}^j\ell_{s}^k 
  + \ell^i_r(2,2,0)|_i
  -\sum_{n=1}^{N}(\frac{1}{2}+\lambda_n)^2\ell_r^iW^{(n)}_i\big) \Big)
\end{multline}
where
\begin{equation}
  V = \frac{1}{6} d_{ijk} a^i a^j a^k .
  \label{60}
\end{equation}

As discussed in~\cite{Lukas:1998hk}, the general form of each $D$-term in the low energy four-dimensional theory is the
sum of 1) the moduli dependent FI parameter \eqref{59} and 2) terms
quadratic in the four-dimensional scalar fields charged under the associated $U(1)$ gauge symmetry weighted by
their specific charge. For each line bundle $L_{r}$, $r=1,\dots,R$ on the Calabi-Yau threefold, there is an anomalous $U(1)_{r}$ symmetry in the four-dimensional low energy theory on the hidden sector. Written in terms of the simplified notation introduced in \cite{Lukas:1999kt,Anderson:2009nt}, the associated D-term is given by
\begin{equation}
D_{r}=FI_{r}-\sum_{L,\bar{M}}Q_{r}^{L} G_{L \bar{M}}C_{r}^{L} {\bar{C}}_{r}^{\bar{M}}.
\label{56a}
\end{equation}
Here $G_{L \bar{M}}$ is an hermitian metric on the $U(1)_{r}$ reducible space of all charged, mass dimension one, scalar matter fields $C^{L}$, which block diagonalizes into the allowed irreducible representations for the r-th line bundle. The indices $L$ and $M$ each run over the full reducible representation, breaking into a sum of the indices over each irreducible sector--each such sector with a unique charge $Q$.
The metric $G_{L \bar{M}}$ is, in general, a complicated function of the K\"ahler moduli with positive definite eigenvalues. 
Note from \eqref{59} that $FI_{r}$ has mass dimension two--consistent with expression \eqref{56a}. As is well-known, a necessary condition for a static vacuum state of the effective four-dimensional theory to be $N=1$ supersymmetric is that the $D$ term associated with each line bundle $L_{r}$ must identically vanish. Generically, this will be the case if
\begin{equation}
\sum_{L,\bar{M}}Q_{r}^{L} G_{L \bar{M}}C_{r}^{L} {\bar{C}}_{r}^{\bar{M}}= FI_{r} \ .
\label{burt1}
\end{equation}
These $C_{r}^{L}$ scalars break into two distinct types-- 1) those that transform only under the Abelian group $U(1)_{r}$ and 2) those which, in addition, transform non-trivially under the non-Abelian gauge factor of the hidden sector low energy theory. This second type of scalar field will also appear in the $D$-term associated with the non-Abelian group--which cannot contain a FI term. Hence, the demand that the vacuum be supersymmetric generically sets their vacuum expectation values to zero. It follow that one can, henceforth, ignore such fields and restrict the scalars in \eqref{burt1} to those that transform under the Abelian $U(1)_{r}$ symmetry only. 

In the weakly coupled heterotic case discussed in \cite{Braun:2013wr}, it was assumed, for simplicity, that the vacuum expectation values $\langle
C_{r}^{L}\rangle$ {\it all} vanish, even for the scalars not transforming under the low energy non-Abelian gauge factor. In that case, each $D_{r}$ will vanish if and only if $FI_{r}=0$. This restriction puts very strong constraints on the choice of the hidden sector vector bundle. Be that as it may, the assumption that all $\langle
C_{r}^{L}\rangle$ vanish and that $FI_{r}=0$ remains a valid constraint for strongly coupled vacua. However, in the strongly coupled case we are now considering, an {\it alternative} set of constraints can be also be adopted. That is,  one can assume that the scalar fields that only transform under the low energy $U(1)_{r}$ groups are, in general, {\it non-vanishing} and that each $D_{r}$ is set to zero by the associated vacuum expectation values $\langle C_{r}^{L} \rangle$ becoming non-zero. For this to be the case, it is essential to specify the hidden sector vector bundle and to compute the pure $U(1)_{r}$ low energy scalar fields  $C_{r}^{L}$ and their associated charges $Q_{r}^{L}$. This is essential because, should the $U(1)_{r}$ charge be positive, then the associated $D_{r}$ term can vanish if and only if $FI_{r}>0$. On the other hand, if the associated charge is negative, then $D_{r}$ can vanish if and only if $FI_{r}<0$. That is, the condition one needs to impose on the Fayet-Iliopoulos terms will depend on the sign of the charges in the scalar spectrum.

\subsection{A Specific Class of Examples}

The constraint equations listed above are technically rather complicated. Therefore, as we did when discussing poly-stability in subsection 3.1, we now analyze the constraint equations within the context of a specific class of $N=1$ supersymmetric hidden sector vector bundles. 
To do this, one must specify the non-Abelian bundle ${\cal{V}}_{N}$ with structure group $SU(N)$, the
number of line bundles $L_r$ and their exact embeddings into the
hidden $E_8$ vector bundle. We will, henceforth, consider hidden sector bundles that may, or may not, contain a non-Abelian factor and, for simplicity, are restricted to contain at most a single line bundle
\begin{equation}
  L = \Ocal_X(\ell^1,\ell^2,\ell^3)
  \label{70}
\end{equation}
where
\begin{equation}
  \ell^1,\ell^2,\ell^3 \in \Z
  , \quad 
  (\ell^1+\ell^2) \tmod 3 = 0 \ .
  \label{71}
\end{equation}
In this case, there is only a single $a_{r}$ coefficient--which we denote simply by $a$.
In addition, one must specify the
number of five-branes in the bulk space. Again, for simplicity, we assume
that there is only one five-brane in this example. It then follows
from \eqref{33}, \eqref{68}, \eqref{69} and \eqref{59}that the
constraints for this restricted class of examples are given by
\begin{align}
 W_i= 
  \big(\tfrac{4}{3},\tfrac{7}{3},-4\big)\big|_i
  -d_{ijk}c_{N}^{jk}+ad_{ijk} \ell^j \ell^k 
  \;& \geq 0
  , \quad i=1,2,3 ,   \label{77}
  \\[1ex]
 d_{ijk} a^i a^j a^k
  - 3 \epsilon_S' \frac{\Rhat}{V^{1/3}} \Big(
  -\big(\tfrac83 a^1 + \tfrac53 a^2 + 4 a^3\big)  \nonumber
  \qquad \\ 
  \qquad+ 2(a^1+a^2) -\big(\tfrac{1}{2}-\lambda\big)^2 a^i W_i\Big) 
  \;& > 0  \label{78}
  \\[1ex]
    d_{ijk} a^i a^j a^k- 3  \epsilon_S' \frac{\Rhat}{V^{1/3}}
    \Big(-d_{ijk}a^{i}c_{N}^{jk}+
   a d_{ijk}a^i \ell^j \ell^k  \nonumber
  \qquad \\
    + 2(a^1+a^2) -(\frac{1}{2}+\lambda)^2 a^i W_i\Big) 
    \;& > 0 \label{79}
   \\[1ex]
    d_{ijk}\ell^ia^ja^k 
    - \epsilon_S' \frac{\Rhat}{V^{1/3}}
    \Big(-d_{ijk}\ell^{i}c_{N}^{jk}+ ad_{ijk}\ell^i \ell^j\ell^k   \nonumber
    \qquad \\
    + \ell^i(2,2,0)|_i
    -(\frac{1}{2}+\lambda)^2\ell^iW_i\Big) 
  \;&  \lessgtr 0 ~,=0  
  \label{80} 
 \end{align}
where $\Rhat$ is an independent modulus and $V$ satisfies relation
\eqref{60}. Note that the expression on the left-hand side of \eqref{80} is a) $> 0$ or $< 0$ if one assumes that some $\langle \phi_{\alpha} \rangle \neq 0$  and that the associated scalar charge $q_{\alpha}$ is positive or negative respectively or b) $=0$ if, alternatively, one assumes that all $\langle \phi_{\alpha} \rangle =0$. 

To proceed, one must specify the the coefficient $a$, as well as the coefficients $c_{N}^{jk}$ of the second Chern class of ${\cal{V}}_{N}$. We begin with the coefficient $a$.
Recall from \eqref{26} that
\begin{equation}
  a=\frac{1}{4\cdot 30} \tr_{E_8} Q^{2} \ ,
  \label{bro1}
\end{equation}
where $Q$ is the generator of the $U(1)$ structure group of the line bundle. 
Hence, the value of coefficient $a$ will depend entirely on the explicit embedding of this $U(1)$ into
the $\Rep{248}$ representation of the hidden sector $E_8$. Here, we will present one explicit example of such an embedding, although, as will discussed elsewhere, this specific type of embedding is not unique. First, assume that the hidden sector gauge bundle is the Whitney sum of a non-Abelian bundle ${\cal{V}}_{N}$ and a line bundle $L$. Choosing the structure group of ${\cal{V}}_{N}$ to be $SU(N)$, embed
\begin{equation}
SU(N) \times U(1) \subset SU(N+1) \ .
\label{bro2}
\end{equation}
The structure group of the line bundle $L$ is identified with the specific $U(1)$ generator in $SU(N+1)$ which commutes with the generators of the chosen $SU(N)$. In the fundamental representation of $SU(N+1)$, the generator of $U(1)$ can then be written as
\begin{equation}
{\rm diag}\big(\underbrace{1,\dots,1}_{N},-N\big) \ .
\label{bro3}
\end{equation}
This specifies the exact embedding of $U(1)$ into $SU(N+1)$. Now choose $SU(N+1)$ to be a factor of a maximal subgroup of $E_{8}$. The decomposition of the $\Rep{248}$ of $E_{8}$ with respect to this maximal subgroup, together with \eqref{bro3}, then determines the generator $Q$.

This is most easily explained by giving a simple explicit example. Let us assume there is {\it no} non-Abelian bundle--only a single line bundle. That is, 
\begin{equation}
V^{(2)}=L \ .
\label{bro4}
\end{equation}
The explicit embedding of $L$ into $E_8$ is chosen as
follows. First, recall that
\begin{equation}
  SU(2) \times E_7 \subset E_8
  \label{72}
\end{equation}
is a maximal subgroup. With respect to $SU(2) \times E_7$, the
$\Rep{248}$ representation of $E_8$ decomposes as
\begin{equation}
  \Rep{248} \longrightarrow 
  (\Rep{1}, \Rep{133}) \oplus (\Rep{2}, \Rep{56}) \oplus (\Rep{3}, \Rep{1}) .
  \label{73}
\end{equation}
Now choose the generator of the $U(1)$ structure group in the fundamental representation of $SU(2)$ to be $(1,-1)$. It follows that
under $SU(2) \rightarrow U(1)$ 
\begin{equation}
  \Rep{2} \longrightarrow 1 \oplus -1 
  \label{74}
\end{equation}
and, hence, under $U(1) \times E_7$
\begin{equation}
  \Rep{248} \longrightarrow 
  (0, \Rep{133}) \oplus 
  \Big( (1, \Rep{56}) \oplus (-1, \Rep{56})\Big) \oplus 
  \Big( (2, \Rep{1}) \oplus (0, \Rep{1}) \oplus (-2, \Rep{1}) \Big) .
\label{75}
\end{equation}
The generator $Q$ of this embedding of the line bundle can be read off
from expression \eqref{75}. Inserting this into \eqref{bro1}, we find that
\begin{equation}
  a=1 .
  \label{76}
\end{equation}
For the choice of a non-Abelian bundle ${\cal{V}}_{N}$ with structure group $SU(N)$, similar calculations give
\begin{equation}
N=2 \Rightarrow a=3, \quad N=3 \Rightarrow a=6, \quad N=4 \Rightarrow a=10 , \quad N=5 \Rightarrow a=15
\label{bro5}
\end{equation}

As discussed previously, one may or may not include a non-Abelian factor in the hidden sector vector bundle. If a non-Abelian factor ${\cal{V}}_{N}$ is to be included, one must specify it exactly. Generically, there are many possibilities for such a bundle.
As an explicit example, let us choose this to be precisely the same $SU(4)$ bundle as in the observable sector described in Subsection 2.1.2. Doing this greatly simplifies the analysis since this ${\cal{V}}_{4}$ bundle is slope-stable with vanishing slope in the same region of K\"ahler moduli space as the observable sector bundle $V^{(1)}$--that is, when the inequalities \eqref{51} are satisfied. 
Since $N=4$, it follows from \eqref{bro5} that coefficient 
\begin{equation}
a=10
\label{F1}
\end{equation}
and, since ${\cal{V}}_{4}$ is identical to $V^{(1)}$, it follows from \eqref{14} that
\begin{equation}
  c_2({\cal{V}}_{4}) = 
  \frac{1}{v^{2/3}} \big(
  \omega_1 \wedge \omega_1+4~\omega_2 \wedge \omega_2
  + 4~\omega_1 \wedge \omega_2 
  \big).
\label{bro6}
\end{equation}
Hence, from \eqref{tase1} the only non-vanishing $c_{N}^{jk}$ coefficients are
\begin{equation}
c_{4}^{11}=1, \quad c_{4}^{22}=4, \quad c_{4}^{12}=c_{4}^{21}=2 \ .
\label{bro7}
\end{equation}
Inserting these coefficients, along with a=10, into \eqref{77},\eqref{78},\eqref{79},\eqref{80} give the appropriate constraint equations for this class of vacua. As discussed above, if one assumes the VEVs of all scalar fields vanish, then the left-hand side of \eqref{80} must be zero. However, if not all $\langle \phi_{\alpha}\rangle$ vanish, then to determine whether the left-hand side of the FI inequality \eqref{78} should be $> 0$ or $< 0$ depends on the sign of the charge of the associated low energy $U(1)$ charged scalars. Since the charge can be different for different choices of the hidden sector bundle, this can only be determined within the context of an explicit example. This will be presented elsewhere.

Of course, these constraints have to be solved
simultaneously with the condition \eqref{51} for the slope-stability
of both the observable and hidden sector non-Abelian vector bundles; that is
\begin{equation}\label{VisStab}
  \begin{gathered}
    \left(
      a^1
      < 
      a^2
      \leq 
      \sqrt{\tfrac{5}{2}} a^1
      \quad\text{and}\quad
      a^3
      <
      \frac{
        -(a^1)^2-3 a^1 a^2+ (a^2)^2
      }{
        6 a^1-6 a^2
      } 
    \right)
    \quad\text{or}\\
    \left(
      \sqrt{\tfrac{5}{2}} a^1
      <
      a^2
      <
      2 a^1
      \quad\text{and}\quad
      \frac{
        2(a^2)^2-5 (a^1)^2
      }{
        30 a^1-12 a^2
      }
      <
      a^3
      <
      \frac{
        -(a^1)^2-3 a^1 a^2+ (a^2)^2
      }{
        6 a^1-6 a^2
      }
    \right) 
  \end{gathered}
\end{equation}

Finally, it is essential to implement equations \eqref{45B} and
\eqref{45C} for the validity of the linear approximation. These
equations depend sensitively on the sign of each component of
$\beta_i^{(0)}$, the value of $\beta_i^{(1)}=W_i$ and the five-brane
location $z_1$. For the specific class of models presented in this section, \eqref{45B} and \eqref{45C} become
\begin{equation}
 2\epsilon_S'\frac{\Rhat}{V^{1/3}}
  \left|
    \beta_i^{(0)} \big(z-\tfrac{1}{2}\big)
    -\frac{1}{2}W_i(\tfrac{1}{2}-\lambda)^2
  \right|
  \ll 
  \left| d_{ijk} a^j a^k \right|
  , \quad z \in [0, \lambda + \tfrac{1}{2}] \label{83}
  \end{equation}
  and
\begin{equation}
2\epsilon_S'\frac{\Rhat}{V^{1/3}}
  \left|
    (\beta_i^{(0)}+W_i)
    \big(z-\tfrac{1}{2}\big)
    -\frac{1}{2}W_i(\tfrac{1}{2}+\lambda)^2
  \right| 
  \ll 
  \left| d_{ijk} a^j a^k \right|
  , \quad z \in [\lambda + \tfrac{1}{2},1] \ .
  \label{84}
\end{equation}
where, as defined in \eqref{pink2}, $\lambda=z_{1}-\frac{1}{2}$.
%
%
%
%
%

\section*{Acknowledgments}
The author would like to thank Yang-Hui He and Rehan Deen for many informative conversations. Burt Ovrut is supported in part by DOE No. DE-SC0007901 and SAS Account 020-0188-2-010202-6603-0338.

\renewcommand{\refname}{Bibliography}
\addcontentsline{toc}{section}{Bibliography} 
\bibliographystyle{unsrt} 
\bibliography{BIBsNew}

\begin{thebibliography}{10}

\bibitem{Gross:1985fr}
David~J. Gross, Jeffrey~A. Harvey, Emil~J. Martinec, and Ryan Rohm.
\newblock Heterotic string theory. 1. the free heterotic string.
\newblock {\em Nucl. Phys.}, B256:253, 1985.

\bibitem{Gross:1985rr}
David~J. Gross, Jeffrey~A. Harvey, Emil~J. Martinec, and Ryan Rohm.
\newblock Heterotic string theory. 2. the interacting heterotic string.
\newblock {\em Nucl. Phys.}, B267:75, 1986.

\bibitem{Witten:1996mz}
Edward Witten.
\newblock {Strong coupling expansion of Calabi-Yau compactification}.
\newblock {\em Nucl.Phys.}, B471:135--158, 1996.

\bibitem{Horava:1995qa}
Petr Horava and Edward Witten.
\newblock Heterotic and type {I} string dynamics from eleven dimensions.
\newblock {\em Nucl. Phys.}, B460:506--524, 1996.

\bibitem{Horava:1996ma}
Petr Horava and Edward Witten.
\newblock {Eleven-dimensional supergravity on a manifold with boundary}.
\newblock {\em Nucl.Phys.}, B475:94--114, 1996.

\bibitem{Greene:1986ar}
Brian~R. Greene, Kelley~H. Kirklin, Paul~J. Miron, and Graham~G. Ross.
\newblock A superstring inspired standard model.
\newblock {\em Phys. Lett.}, B180:69, 1986.

\bibitem{Greene:1986bm}
Brian~R. Greene, Kelley~H. Kirklin, Paul~J. Miron, and Graham~G. Ross.
\newblock A three generation superstring model. 1. compactification and
  discrete symmetries.
\newblock {\em Nucl. Phys.}, B278:667, 1986.

\bibitem{Greene:1986jb}
Brian~R. Greene, Kelley~H. Kirklin, Paul~J. Miron, and Graham~G. Ross.
\newblock A three generation superstring model. 2. symmetry breaking and the
  low-energy theory.
\newblock {\em Nucl. Phys.}, B292:606, 1987.

\bibitem{Matsuoka:1986vg}
Takeo Matsuoka and Daijiro Suematsu.
\newblock Realistic models from the {$E_8\times E_8'$} superstring theory.
\newblock {\em Prog. Theor. Phys.}, 76:886, 1986.

\bibitem{Greene:1987xh}
Brian~R. Greene, K.~H. Kirklin, P.~J. Miron, and Graham~G. Ross.
\newblock {$27^3$} {Yukawa} couplings for a three generation superstring model.
\newblock {\em Phys. Lett.}, B192:111, 1987.

\bibitem{Anderson:2011cza}
Lara~B. Anderson, James Gray, Andre Lukas, and Burt Ovrut.
\newblock {Stabilizing All Geometric Moduli in Heterotic Calabi-Yau Vacua}.
\newblock {\em Phys. Rev.}, D83:106011, 2011.

\bibitem{Braun:2005nv}
Volker Braun, Yang-Hui He, Burt~A. Ovrut, and Tony Pantev.
\newblock {The Exact MSSM spectrum from string theory}.
\newblock {\em JHEP}, 0605:043, 2006.

\bibitem{Braun:2004xv}
Volker Braun, Burt~A. Ovrut, Tony Pantev, and Rene Reinbacher.
\newblock {Elliptic Calabi-Yau threefolds with Z(3) x Z(3) Wilson lines}.
\newblock {\em JHEP}, 0412:062, 2004.

\bibitem{Braun:2005zv}
Volker Braun, Yang-Hui He, Burt~A. Ovrut, and Tony Pantev.
\newblock {Vector Bundle Extensions, Sheaf Cohomology, and the Heterotic
  Standard Model}.
\newblock {\em Adv. Theor. Math. Phys.}, 10:4, 2006.

\bibitem{Braun:2006ae}
Volker Braun, Yang-Hui He, and Burt~A. Ovrut.
\newblock {Stability of the minimal heterotic standard model bundle}.
\newblock {\em JHEP}, 0606:032, 2006.

\bibitem{Braun:2006me}
Volker Braun, Yang-Hui He, and Burt~A. Ovrut.
\newblock {Yukawa couplings in heterotic standard models}.
\newblock {\em JHEP}, 0604:019, 2006.

\bibitem{Barger:2008wn}
Vernon Barger, Pavel Fileviez~Perez, and Sogee Spinner.
\newblock {Minimal gauged U(1)(B-L) model with spontaneous R-parity violation}.
\newblock {\em Phys.Rev.Lett.}, 102:181802, 2009.

\bibitem{FileviezPerez:2009gr}
Pavel Fileviez~Perez and Sogee Spinner.
\newblock {Spontaneous R-Parity Breaking in SUSY Models}.
\newblock {\em Phys.Rev.}, D80:015004, 2009.

\bibitem{Ambroso:2009jd}
Michael Ambroso and Burt Ovrut.
\newblock {The B-L/Electroweak Hierarchy in Heterotic String and M-Theory}.
\newblock {\em JHEP}, 0910:011, 2009.

\bibitem{Ambroso:2009sc}
Michael Ambroso and Burt~A. Ovrut.
\newblock {The B-L/Electroweak Hierarchy in Smooth Heterotic
  Compactifications}.
\newblock {\em Int.J.Mod.Phys.}, A25:2631--2677, 2010.

\bibitem{Ambroso:2010pe}
Michael Ambroso and Burt~A. Ovrut.
\newblock {The Mass Spectra, Hierarchy and Cosmology of B-L MSSM Heterotic
  Compactifications}.
\newblock {\em Int.J.Mod.Phys.}, A26:1569--1627, 2011.

\bibitem{Ovrut:2012wg}
Burt~A. Ovrut, Austin Purves, and Sogee Spinner.
\newblock {Wilson Lines and a Canonical Basis of SU(4) Heterotic Standard
  Models}.
\newblock {\em JHEP}, 1211:026, 2012.

\bibitem{Ovrut:2015uea}
Burt~A. Ovrut, Austin Purves, and Sogee Spinner.
\newblock {The minimal SUSY $B − L$ model: from the unification scale to the
  LHC}.
\newblock {\em JHEP}, 06:182, 2015.

\bibitem{FileviezPerez:2009bw}
Pavel Fileviez~Perez and Sogee Spinner.
\newblock {TeV Scale Spontaneous R-Parity Violation}.
\newblock {\em AIP Conf.Proc.}, 1200:529--532, 2010.

\bibitem{FileviezPerez:2012mj}
Pavel Fileviez~Perez and Sogee Spinner.
\newblock {The Minimal Theory for R-parity Violation at the LHC}.
\newblock {\em JHEP}, 1204:118, 2012.

\bibitem{Brelidze:2010hf}
Tamaz Brelidze and Burt~A. Ovrut.
\newblock {B-L Cosmic Strings in Heterotic Standard Models}.
\newblock {\em JHEP}, 1007:077, 2010.

\bibitem{Battarra:2014tga}
Lorenzo Battarra, Michael Koehn, Jean-Luc Lehners, and Burt~A. Ovrut.
\newblock {Cosmological Perturbations Through a Non-Singular
  Ghost-Condensate/Galileon Bounce}.
\newblock {\em JCAP}, 1407:007, 2014.

\bibitem{Koehn:2013upa}
Michael Koehn, Jean-Luc Lehners, and Burt~A. Ovrut.
\newblock {Cosmological super-bounce}.
\newblock {\em Phys. Rev.}, D90(2):025005, 2014.

\bibitem{Koehn:2012ar}
Michael Koehn, Jean-Luc Lehners, and Burt~A. Ovrut.
\newblock {Higher-Derivative Chiral Superfield Actions Coupled to N=1
  Supergravity}.
\newblock {\em Phys. Rev.}, D86:085019, 2012.

\bibitem{Marshall:2014kea}
Zachary Marshall, Burt~A. Ovrut, Austin Purves, and Sogee Spinner.
\newblock {Spontaneous $R$-Parity Breaking, Stop LSP Decays and the Neutrino
  Mass Hierarchy}.
\newblock {\em Phys. Lett.}, B732:325--329, 2014.

\bibitem{Marshall:2014cwa}
Zachary Marshall, Burt~A. Ovrut, Austin Purves, and Sogee Spinner.
\newblock {LSP Squark Decays at the LHC and the Neutrino Mass Hierarchy}.
\newblock {\em Phys. Rev.}, D90(1):015034, 2014.

\bibitem{Dumitru:2018jyb}
Sebastian Dumitru, Burt~A. Ovrut, and Austin Purves.
\newblock {The $R$-parity Violating Decays of Charginos and Neutralinos in the
  B-L MSSM}.
\newblock {\em JHEP}, 02:124, 2019.

\bibitem{Dumitru:2018nct}
Sebastian Dumitru, Burt~A. Ovrut, and Austin Purves.
\newblock {$R$-parity Violating Decays of Wino Chargino and Wino Neutralino
  LSPs and NLSPs at the LHC}.
\newblock {\em JHEP}, 06:100, 2019.

\bibitem{Dumitru:2019cgf}
Sebastian Dumitru, Christian Herwig, and Burt~A. Ovrut.
\newblock {$R$-parity Violating Decays of Bino Neutralino LSPs at the LHC}.
\newblock {\em JHEP}, 12:042, 2019.

\bibitem{Candelas:2007ac}
Philip Candelas, Xenia de~la Ossa, Yang-Hui He, and Balazs Szendroi.
\newblock {Triadophilia: A Special Corner in the Landscape}.
\newblock {\em Adv.Theor.Math.Phys.}, 12:429--473, 2008.

\bibitem{Anderson:2011ns}
Lara~B. Anderson, James Gray, Andre Lukas, and Eran Palti.
\newblock {Two Hundred Heterotic Standard Models on Smooth Calabi-Yau
  Threefolds}.
\newblock {\em Phys.Rev.}, D84:106005, 2011.

\bibitem{Anderson:2012yf}
Lara~B. Anderson, James Gray, Andre Lukas, and Eran Palti.
\newblock {Heterotic Line Bundle Standard Models}.
\newblock {\em JHEP}, 1206:113, 2012.

\bibitem{Anderson:2011vy}
Lara~B. Anderson, James Gray, Andre Lukas, and Eran Palti.
\newblock {Heterotic standard models from smooth Calabi-Yau three-folds}.
\newblock {\em PoS}, CORFU2011:096, 2011.

\bibitem{Kachru:2003aw}
Shamit Kachru, Renata Kallosh, Andrei~D. Linde, and Sandip~P. Trivedi.
\newblock {De Sitter vacua in string theory}.
\newblock {\em Phys.Rev.}, D68:046005, 2003.

\bibitem{Gray:2007zza}
James Gray, Andre Lukas, and Burt Ovrut.
\newblock {Perturbative anti-brane potentials in heterotic M-theory}.
\newblock {\em Phys.Rev.}, D76:066007, 2007.

\bibitem{Gray:2007qy}
James Gray, Andre Lukas, and Burt Ovrut.
\newblock {Flux, gaugino condensation and anti-branes in heterotic M-theory}.
\newblock {\em Phys.Rev.}, D76:126012, 2007.

\bibitem{Braun:2006th}
Volker Braun and Burt~A. Ovrut.
\newblock {Stabilizing moduli with a positive cosmological constant in
  heterotic M-theory}.
\newblock {\em JHEP}, 0607:035, 2006.

\bibitem{MR522939}
F.~A. Bogomolov.
\newblock Holomorphic tensors and vector bundles on projective manifolds.
\newblock {\em Izv. Akad. Nauk SSSR Ser. Mat.}, 42(6):1227--1287, 1439, 1978.

\bibitem{Douglas:2006jp}
Michael~R. Douglas, Rene Reinbacher, and Shing-Tung Yau.
\newblock {Branes, bundles and attractors: Bogomolov and beyond}.
\newblock 2006.

\bibitem{Andreas:2010hv}
Bjorn Andreas and Gottfried Curio.
\newblock {Spectral Bundles and the DRY-Conjecture}.
\newblock {\em J.Geom.Phys.}, 62:800--803, 2012.

\bibitem{Andreas:2011zs}
Bjorn Andreas and Gottfried Curio.
\newblock {On the Existence of Stable bundles with prescribed Chern classes on
  Calabi-Yau threefolds}.
\newblock 2011.

\bibitem{Braun:2013wr}
Volker Braun, Yang-Hui He, and Burt~A. Ovrut.
\newblock {Supersymmetric Hidden Sectors for Heterotic Standard Models}.
\newblock {\em JHEP}, 1309:008, 2013.

\bibitem{Blumenhagen:2005ga}
Ralph Blumenhagen, Gabriele Honecker, and Timo Weigand.
\newblock {Loop-corrected compactifications of the heterotic string with line
  bundles}.
\newblock {\em JHEP}, 0506:020, 2005.

\bibitem{Blumenhagen:2005zg}
Ralph Blumenhagen, Gabriele Honecker, and Timo Weigand.
\newblock {Non-Abelian brane worlds: The Heterotic string story}.
\newblock {\em JHEP}, 0510:086, 2005.

\bibitem{Blumenhagen:2006ux}
Ralph Blumenhagen, Sebastian Moster, and Timo Weigand.
\newblock {Heterotic GUT and standard model vacua from simply connected
  Calabi-Yau manifolds}.
\newblock {\em Nucl.Phys.}, B751:186--221, 2006.

\bibitem{Weigand:2006yj}
T.~Weigand.
\newblock {Compactifications of the heterotic string with unitary bundles}.
\newblock {\em Fortsch.Phys.}, 54:963--1077, 2006.

\bibitem{Blumenhagen:2006wj}
Ralph Blumenhagen, Sebastian Moster, Rene Reinbacher, and Timo Weigand.
\newblock {Massless Spectra of Three Generation U(N) Heterotic String Vacua}.
\newblock {\em JHEP}, 0705:041, 2007.

\bibitem{Blumenhagen:2005pm}
Ralph Blumenhagen, Gabriele Honecker, and Timo Weigand.
\newblock {Supersymmetric (non-)Abelian bundles in the Type I and SO(32)
  heterotic string}.
\newblock {\em JHEP}, 0508:009, 2005.

\bibitem{Lukas:1998hk}
Andre Lukas, Burt~A. Ovrut, and Daniel Waldram.
\newblock {Nonstandard embedding and five-branes in heterotic M theory}.
\newblock {\em Phys.Rev.}, D59:106005, 1999.

\bibitem{Donagi:1999jp}
Ron Donagi, Burt~A. Ovrut, and Daniel Waldram.
\newblock {Moduli spaces of five-branes on elliptic Calabi-Yau threefolds}.
\newblock {\em JHEP}, 11:030, 1999.

\bibitem{Lukas:1999kt}
Andre Lukas, Burt~A. Ovrut, and Daniel Waldram.
\newblock {Five-branes and supersymmetry breaking in M theory}.
\newblock {\em JHEP}, 9904:009, 1999.

\bibitem{Lukas:1997fg}
Andre Lukas, Burt~A. Ovrut, and Daniel Waldram.
\newblock On the four-dimensional effective action of strongly coupled
  heterotic string theory.
\newblock {\em Nucl. Phys.}, B532:43--82, 1998.

\bibitem{Anderson:2009nt}
Lara~B. Anderson, James Gray, Andre Lukas, and Burt Ovrut.
\newblock {Stability Walls in Heterotic Theories}.
\newblock {\em JHEP}, 0909:026, 2009.

\bibitem{MR923487}
Chad Schoen.
\newblock On fiber products of rational elliptic surfaces with section.
\newblock {\em Math. Z.}, 197(2):177--199, 1988.

\bibitem{Lukas:1998tt}
Andre Lukas, Burt~A. Ovrut, K.S. Stelle, and Daniel Waldram.
\newblock {Heterotic M theory in five-dimensions}.
\newblock {\em Nucl.Phys.}, B552:246--290, 1999.

\bibitem{Donagi:1998xe}
Ron Donagi, Andre Lukas, Burt~A. Ovrut, and Daniel Waldram.
\newblock {Nonperturbative vacua and particle physics in M theory}.
\newblock {\em JHEP}, 9905:018, 1999.

\bibitem{Lukas:1998yy}
Andre Lukas, Burt~A. Ovrut, K.S. Stelle, and Daniel Waldram.
\newblock {The Universe as a domain wall}.
\newblock {\em Phys.Rev.}, D59:086001, 1999.

\bibitem{Donagi:1999gc}
Ron Donagi, Andre Lukas, Burt~A. Ovrut, and Daniel Waldram.
\newblock {Holomorphic vector bundles and nonperturbative vacua in M theory}.
\newblock {\em JHEP}, 9906:034, 1999.

\bibitem{MR88i:58154}
K.~Uhlenbeck and S.-T. Yau.
\newblock On the existence of hermitian-{Y}ang-{M}ills connections in stable
  vector bundles.
\newblock {\em Comm. Pure Appl. Math.}, 39(S, suppl.):S257--S293, 1986.
\newblock Frontiers of the mathematical sciences: 1985 (New York, 1985).

\bibitem{MR86h:58038}
S.~K. Donaldson.
\newblock Anti self-dual {Y}ang-{M}ills connections over complex algebraic
  surfaces and stable vector bundles.
\newblock {\em Proc. London Math. Soc. (3)}, 50(1):1--26, 1985.

\bibitem{Green:1987mn}
Michael~B. Green, J.~H. Schwarz, and Edward Witten.
\newblock Superstring theory. vol. 2: Loop amplitudes, anomalies and
  phenomenology.
\newblock Cambridge, Uk: Univ. Pr. ( 1987) 596 P. ( Cambridge Monographs On
  Mathematical Physics).

\bibitem{Banks:1996ss}
Tom Banks and Michael Dine.
\newblock {Couplings and scales in strongly coupled heterotic string theory}.
\newblock {\em Nucl.Phys.}, B479:173--196, 1996.

\end{thebibliography}

\end{document}